\documentclass[11pt]{article}
\usepackage[T1]{fontenc}
\usepackage[utf8]{inputenc}
\usepackage{lmodern}
\usepackage[margin=1in]{geometry}
\usepackage{amsmath,amssymb,amsfonts,mathtools}
\usepackage{amsthm}
\usepackage{booktabs,array,longtable,tabularx}
\usepackage{enumitem}
\usepackage{microtype}
\usepackage[round,authoryear]{natbib}
\usepackage[hidelinks]{hyperref}
\usepackage{xurl}
\usepackage{setspace}
\allowdisplaybreaks
\providecommand{\tightlist}{\setlength{\itemsep}{0pt}\setlength{\parskip}{0pt}}

\title{Orthogonal Validation-Augmented Cox Regression with Internally Validated Failure Indicators\\[0.5em]
\large Cross-Fitted Estimation, Risk-Set Linearization, and Validation Design}
\author{Subir Hait\\
Department of Counseling, Educational Psychology and Special Education\\
Michigan State University\\
\texttt{haitsubi@msu.edu}\\
ORCID: 0009-0004-9871-9677}
\date{August 2026}

\begin{document}
\maketitle
\clearpage

\begin{abstract}
Incomplete adjudication and endpoint misclassification arise when gold-standard event classification is available only for an internal validation sample. With accurate follow-up times and covariates but error-prone event indicators, naive Cox regression can be biased and inverse-probability weighting inefficient. We develop orthogonal validation-augmented Cox (OVAC) regression, a cross-fitted augmented Cox estimator for incomplete failure indicators. OVAC replaces the missing event indicator with a cross-fitted augmented pseudo-event and explicitly retains the first-order contribution from estimating the Cox risk-set mean. The same term emerges from projection of the full-data influence function through the validation mechanism, linking the score expansion to the observed-data efficient influence function. The score exhibits exact product-form nuisance drift, yielding double-robust identification, Neyman orthogonality, and root-$n$ inference under product-rate conditions. In 1,000 R replications, full-variance SE/SD ratios were 0.997 and 0.992, with 95\% coverage of 0.955 and 0.948. A 200-replication diagnostic showed that the risk-set term alone was 77.8\% and 81.1\% of the full-SE magnitude, although covariance cancellation made its net SE effect smaller than 0.3\%. In a 250-replication method comparison, OVAC reduced empirical variance by 38.8\% and 37.7\% relative to IPW. OVAC provides machine-learning-compatible Cox estimation with explicit risk-set linearization, calibrated inference, and validation-design guidance.
\end{abstract}

\clearpage

\section{Introduction}

Time-to-event analyses increasingly draw on electronic health records,
administrative databases, disease registries, and surveillance systems.
These sources can provide large cohorts and long follow-up, yet the
event label may be less reliable than the recorded follow-up time or
baseline covariates. A diagnosis code may fail clinical adjudication, a
reported event may later be reclassified, or an end-of-follow-up record
may be difficult to distinguish from the event of interest. When the
termination time is accurate but its event label is uncertain, treating
the error-prone label as a gold-standard failure indicator can distort
Cox regression because the event indicator determines which risk-set
contrasts enter the partial score \citep{cox1972,andersen1982}.

Internal validation provides a natural design response. In phase I,
every subject contributes accurately observed follow-up time, baseline
covariates, an error-prone event indicator, and possibly other
inexpensive auxiliary information. In phase II, a subset undergoes chart
review, expert adjudication, linkage to a higher-quality source, or
another gold-standard assessment that reveals the true failure
indicator. Validation is often deliberately outcome-dependent with
respect to phase-I information: apparent events, uncertain records,
particular sites, or high-risk profiles may be oversampled.
Complete-case analysis of validated records can therefore be inefficient
and can be biased if the validation mechanism is ignored \citep{edwards2013,tao2021}.

The statistical problem is classical; the computational setting is not.
Modern applications may contain many phase-I predictors of adjudicated
status, making flexible machine learning attractive for estimating the
conditional event probability while the target remains a low-dimensional
Cox regression coefficient. This combination motivates orthogonal scores
and sample splitting: nuisance functions can be estimated adaptively
while first-order inference for the target parameter is insulated from
small nuisance-estimation errors \citep{chernozhukov2018,kennedy2016}.

\subsection{Prior work on missing and misclassified failure
indicators}

The relevant novelty comparison is the direct literature on missing or
misclassified failure indicators, not only the broader measurement-error
literature. Methods that recover information from subjects whose failure
indicators are unavailable predate modern double machine learning. \citet{vanderlaan1998} developed efficient estimation for
right-censored data with missing failure indicators. \citet{mckeague1998} studied product-limit estimation and Cox regression
with missing cause-of-failure information. \citet{gijbels2007}
introduced estimating functions for Cox regression with missing failure
indicators and constructed an adaptive estimator attaining the minimum
variance-covariance bound within their class.

Incomplete endpoint adjudication provides a closely related formulation.
\citet{cook2004} analyzed time-to-event data with incomplete
event adjudication and used estimated probabilities that unadjudicated
events were true events. \citet{magaret2008} developed validation-subset
methods for discrete proportional-hazards models with mismeasured
outcomes. \citet{brownstein2015} compared approaches for Cox
regression with missing failure indicators, including multiple
imputation in the OPPERA study. \citet{ni2017} used internal
validation and multiple imputation to correct hazard-ratio estimates for
outcome misclassification.

Most directly, \citet{liu2010} developed regression-imputation and
augmented inverse-probability-weighted estimators for Cox regression
when failure indicators are missing at random. Their augmented event
factor can be written as

\[m(V) + \frac{R}{\pi(V)}\left( \Delta - m(V) \right),\]

which is algebraically the same missing-data augmentation used below.
Their work established robustness and asymptotic normality under
parametric and nonparametric nuisance estimation. Accordingly, OVAC does
not claim the augmented pseudo-event itself as a new estimator.

A broader survival-measurement-error literature treats harder problems
in which event times, exposures, covariates, or causes of failure are
also error-prone. Examples include augmented estimation for missing
covariates \citep{wang2001}, corrected-score and internal-validation
methods for covariate error \citep{zucker2008,zucker2019}, cause-specific survival models with misclassified causes \citep{vanrompaye2012}, and SIMEX, regression calibration, generalized
raking, quasi-likelihood, and multiple imputation for broader
time-to-event measurement error \citep{boe2021,giganti2020,oh2018,oh2021a,oh2021b}. Recent work also treats outcome
measurement error in risk and survival functions using validation
information \citep{edwards2026}. Two-phase sampling theory provides a
general foundation for weighted and augmented semiparametric estimation
\citep{breslow2007,robins1994,saegusa2013,tao2021}. More recent work uses influence functions to
construct adaptive multi-wave validation designs \citep{han2021,shepherd2023}.

\subsection{Contribution and novelty
boundary}

The scope is intentionally narrow. The observed follow-up time \(U\) and
regression covariates \(X\) are assumed accurate for the full cohort, while
the gold-standard failure indicator \(\Delta\) is observed only for validated
subjects. An error-prone indicator \(\widetilde{\Delta}\) and other phase-I auxiliary
variables may be observed for everyone. Thus uncertainty concerns
whether the recorded termination at \(U\) is a true event, not whether an
event occurred at an unobserved earlier time. If event time itself is
error-prone, the risk sets are also contaminated and the method
developed here is insufficient.

That restriction creates the key structural simplification: Cox risk
sets are fully observed. Conditional on those risk sets, the full-data
partial score is linear in \(\Delta\). Missing-data augmentation can therefore be
inserted directly into the event contribution without modeling a latent
risk-set process.

Our contribution is not a new augmented event factor; it is a
first-order theory and implementation for the cross-fitted pseudo-event
Cox score. We make five contributions. First, we derive the
empirical-score linearization and explicitly retain the first-order
contribution from estimation of the Cox risk-set mean. Augmenting the
event indicator does not remove this contribution, and an influence
representation intended to match the pseudo-event score must account for
it. Second, we show that the population drift relative to the full-data
Cox score factorizes exactly as the product of event-bridge and
validation-probability errors, providing a transparent double-robustness
and Neyman-orthogonality argument. Third, we combine this score with
cross-fitting and establish root-\(n\) inference under product-rate
conditions, allowing flexible nuisance learners without Donsker-class
restrictions. Fourth, we identify the observed-data efficient influence
function at the intersection model and show that the score linearization
leads to a practical analytic variance estimator. Fifth, we specialize
influence-function design theory to endpoint validation and obtain a
contrast-specific allocation rule that targets records that are
simultaneously classification-ambiguous and influential for the Cox
coefficient. The numerical decomposition in Section 9 clarifies an
important distinction: the risk-set term is first-order in magnitude,
but its net effect on variance can be small when its own variance is
offset by covariance with the event-score component. That cancellation
is a feature of a data-generating regime, not a reason to omit the term
from the influence representation.

OVAC should therefore be viewed as a cross-fitted,
machine-learning-compatible extension and re-expression of classical
augmented Cox estimation for incomplete failure indicators. With
finite-dimensional nuisance models fitted without sample splitting, the
point-estimating equation reduces to the familiar augmented
inverse-probability form. The methodological contribution is the
explicit first-order treatment of the empirical risk sets, the
orthogonal cross-fitted formulation and product-rate theory, the
corresponding efficient-regime variance construction, and the
validation-design specialization. We do not claim that omitting the
risk-set term must always produce a large or directionally predictable
finite-sample error; instead, the term is required by the score
expansion and efficient influence function, while its net variance
impact depends on the covariance structure of the design.

Section 2 defines the full-data Cox model, validation structure, and
target parameter. Section 3 introduces the OVAC score and its
relationship to classical augmented estimators. Section 4 establishes
identification, double robustness, and orthogonality. Section 5 develops
the cross-fitted large-sample theory, and Section 6 derives the
observed-data efficient influence function and practical variance
estimator. Section 7 treats validation design, Section 8 discusses
interpretation and extensions, Section 9 reports Monte Carlo
calibration, and Section 10 gives implementation guidance. Section 11
discusses implications and concludes, and Section 12 reports acknowledgments
and declarations. Proofs and computational details appear in the appendices.

\section{Statistical framework}

\subsection{Full-data survival
process}

For subject \(i\), let \(T_{i}^{0}\) denote the event time of scientific
interest and \(C_{i}\) a right-censoring time. We observe
\(U_{i} = \min\left( T_{i}^{0},C_{i} \right)\) for every subject and
define the true event indicator
\(\Delta_{i} = I\left( T_{i}^{0} \leq C_{i} \right)\). Let \(X_{i}\) be
a \(p\)-dimensional, time-fixed vector of regression covariates. The
core theory assumes that \(X_{i}\) and \(U_{i}\) are accurately observed
for the entire phase-I cohort. Extensions to external time-dependent
covariates are discussed later.

The target full-data model is the Cox proportional hazards model \citep{cox1972}

\[
\lambda(t \mid X) = \lambda_{0}(t)\exp\left( \beta_{0}^{\top}X \right),\quad\quad 0 \leq t \leq \tau.
\tag{1}\label{eq:1}
\]

where \(\beta_{0} \in \mathbb{R}^{p}\) is the target regression parameter and
\(\lambda_{0}\) is an unspecified baseline hazard. Define the at-risk and counting
processes \(Y_{i}(t)=I(U_{i}\ge t)\) and
\(N_{i}(t)=I(U_{i}\le t,\Delta_{i}=1)\). Under conditional independent
censoring and standard regularity conditions, the partial-likelihood score
identifies \(\beta_{0}\) \citep{andersen1982,fleming1991}.

\[
s^{(k)}(t,\beta) = E\left\lbrack Y(t)\exp\left( \beta^{\top}X \right)X^{\otimes k} \right\rbrack,\quad\quad k = 0,1,2.
\tag{2}\label{eq:2}
\]

\[
\bar{X}(t,\beta) = \frac{s^{(1)}(t,\beta)}{s^{(0)}(t,\beta)}.
\tag{3}\label{eq:3}
\]

\[
v(t,\beta) = \frac{s^{(2)}(t,\beta)}{s^{(0)}(t,\beta)} - \bar{X}(t,\beta)^{\otimes 2}.
\tag{4}\label{eq:4}
\]

For a single-event process with time-fixed covariates, the population
Cox score may be written compactly as

\[
\Psi_{F}(\beta) = E\left\lbrack \Delta\left( X - \bar{X}(U,\beta) \right) \right\rbrack.
\tag{5}\label{eq:5}
\]

The equivalent counting-process form is
\(E\int_{0}^{\tau}\left( X - \bar{X}(t,\beta) \right)dN(t)\). The target
parameter \(\beta_{0}\) is the unique zero of \(\Psi_{F}\) in a
neighborhood of the truth.

\subsection{Error-prone failure indicator and internal
validation}

The true event indicator is not observed for every subject. Instead, all
phase-I records contain an error-prone indicator \(\widetilde{\Delta}\)
and a vector \(A\) of auxiliary variables that can include diagnostic
codes, laboratory features, text-derived confidence scores, site,
calendar time, or other information available before validation. We
collect these variables into
\(V = \left( U,X,\widetilde{\Delta},A \right)\). Let \(R = 1\) indicate
that a gold-standard review is performed, in which case \(\Delta\) is
observed. The observed data are \(O = (V,R,R\Delta)\).

We allow validation to be outcome-dependent with respect to the error-prone
information. For example, records with \(\widetilde{\Delta}=1\) may be
oversampled because apparent events are scientifically important, or records
with an uncertainty score near a decision threshold may be preferentially
adjudicated. The identifying assumption is not simple random validation but
missing at random conditional on \(V\), consistent with standard two-phase
and missing-data formulations \citep{robins1994,breslow2007,tao2021}.

\[
R\bot\bot\Delta \mid V.
\tag{6}\label{eq:6}
\]

\[
\pi_{0}(V) = \Pr(R = 1 \mid V),\quad\quad\varepsilon \leq \pi_{0}(V) \leq 1.
\tag{7}\label{eq:7}
\]

for some \(\varepsilon > 0\). In a designed validation study \(\pi_{0}\)
may be known exactly. In retrospective settings it may be estimated from
observed validation indicators and phase-I variables.

\[
p_{0}(V) = \Pr(\Delta = 1 \mid V).
\tag{8}\label{eq:8}
\]

We call \(p_{0}\) the event-classification bridge because it links the
phase-I record to the gold-standard event status. No
sensitivity/specificity parameterization is required. Differential
misclassification is allowed because \(p_{0}\) may depend on \(X\),
\(U\), \(\widetilde{\Delta}\), and auxiliary variables.

\subsection{Target estimand and
scope}

The target is the same \(\beta_{0}\) that would be estimated if
\(\Delta\) were gold-standard for the entire cohort. Thus the estimand
is defined by the full-data Cox model rather than by the error-prone
indicator. This distinction matters: the method is not attempting to
estimate a hazard model for \(\widetilde{\Delta}\). It uses phase-I
information only to recover the full-data score.

The main theory assumes that U is accurately measured. If the event time
itself is shifted, rounded, interval-censored, or otherwise error-prone,
then Y(t) and the risk sets are also contaminated and the linear
pseudo-event simplification no longer suffices. That broader problem is
important but intentionally outside the main model. Likewise, the core
results do not address error-prone regression covariates, informative
censoring beyond the stated conditioning set, recurrent events, or
competing risks. Section 8 outlines extensions and clarifies which parts
of the theory survive.

\section{Orthogonal validation-augmented Cox
estimation}

\subsection{The augmented
pseudo-event}

For candidate nuisance functions \(p\) and \(\pi\), define the augmented
pseudo-event

\[
D(p,\pi) = p(V) + \frac{R}{\pi(V)}\left( \Delta - p(V) \right).
\tag{9}\label{eq:9}
\]

The pseudo-event has two components. The regression term \(p(V)\) predicts
the true failure indicator for every subject. For validated subjects,
the residual \(\Delta\)-p(V) is returned to the cohort after inverse-probability
scaling. If \(\pi\) is correct, this residual correction removes bias from a
misspecified event bridge. If \(p\) is correct, the residual has conditional
mean zero given \(V\), so consistency does not require a correct
validation-probability model.

\(D(p,\pi)\) is not constrained to lie in \(\lbrack 0,1\rbrack\). That
is not a defect: it is an estimating-equation pseudo-outcome rather than
a literal event probability. Negative values can occur for validated
non-events when \(\pi < 1\). The target score remains well defined
because \(D\) enters linearly as an event contribution.

\subsection{Population and sample
scores}

\[
c_{\beta}(U,X) = X - \bar{X}(U,\beta).
\tag{10}\label{eq:10}
\]

\[
\Psi(\beta;p,\pi) = E\left( D(p,\pi)c_{\beta}(U,X) \right).
\tag{11}\label{eq:11}
\]

If either nuisance component is correct, Section 4 shows that
\(\Psi(\beta;p,\pi) = \Psi_{F}(\beta)\), so \(\beta_{0}\) remains the
unique root.

In a sample of size n, define the ordinary empirical risk-set quantities

\[
S_{n}^{(k)}(t,\beta) = \frac{1}{n}\sum_{i = 1}^{n}Y_{i}(t)\exp\left( \beta^{\top}X_{i} \right)X_{i}^{\otimes k}.
\tag{12}\label{eq:12}
\]

\[
\bar{X}_{n}(t,\beta) = \frac{S_{n}^{(1)}(t,\beta)}{S_{n}^{(0)}(t,\beta)}.
\tag{13}\label{eq:13}
\]

Let \({\widehat{p}}_{- k(i)}\) and \({\widehat{\pi}}_{- k(i)}\) be
nuisance estimates trained without the fold containing observation
\(i\). Define

\[
{\widetilde{D}}_{i} = {\widehat{p}}_{- k(i)}\left( V_{i} \right) + \frac{R_{i}}{{\widehat{\pi}}_{- k(i)}\left( V_{i} \right)}\left( \Delta_{i} - {\widehat{p}}_{- k(i)}\left( V_{i} \right) \right).
\tag{14}\label{eq:14}
\]

\[
U_{n}(\beta) = \frac{1}{n}\sum_{i = 1}^{n}{\widetilde{D}}_{i}\left( X_{i} - \bar{X}_{n}\left( U_{i},\beta \right) \right).
\tag{15}\label{eq:15}
\]

The OVAC estimator \(\widehat{\beta}\) is any consistent root of
\(U_{n}(\beta) = 0\). In practice Newton-Raphson uses the same risk-set
cumulative sums as ordinary Cox regression, replacing the binary event
indicator by \(\widetilde{D}\) only in event contributions and the
observed information.

\subsection{Algorithm 1: cross-fitted
OVAC}

\begin{itemize}
\item
  Choose \(K\) folds. If validation probabilities are known by design,
  set \(\widehat{\pi} = \pi_{0}\) and skip validation-model fitting.
\item
  For each held-out fold, fit \(p(V) = \Pr(\Delta = 1 \mid V)\) using
  validated records in the training folds. Fit
  \(\pi(V) = \Pr(R = 1 \mid V)\) on the training folds if \(\pi_{0}\) is
  not known.
\item
  Predict \(\widehat{p}\) and \(\widehat{\pi}\) for the held-out
  observations, truncate \(\widehat{\pi}\) away from zero if required by
  the design, and compute the pseudo-event \(\widetilde{D}\).
\item
  Using all accurately observed \(U\) and \(X\), form standard Cox risk
  sets and solve \(U_{n}(\beta) = 0\) with \(\widetilde{D}\) in place of
  \(\Delta\).
\item
  For efficient-regime inference, estimate the influence function in
  Section 6 using the cross-fitted nuisance estimates, the risk-set
  means at \(\widehat{\beta}\), and the signed score-linearization
  measure defined in Section 6.4. Use the bootstrap if relying only on
  one side of double robustness or when nuisance estimation is highly
  adaptive and finite-sample regularity is uncertain.
\end{itemize}

Because \(p(V)=\Pr(\Delta=1\mid V)\) is a conditional probability, any probabilistic learner can in
principle be used: logistic regression, splines, boosted trees, random
forests, Super Learner, or a calibrated neural classifier. Cross-fitting
separates nuisance training from score evaluation and prevents the
empirical-process complexity of the learner from entering the
first-order expansion in the usual way \citep{chernozhukov2018,kennedy2016}.

\subsection{Relationship to classical augmented Cox
estimators}

Let \(m(V)\) denote a working model for \(\Pr(\Delta = 1 \mid V)\) and
\(\pi(V)\) the validation probability. The AIPW event contribution
studied in earlier missing-failure-indicator work can be rearranged as

\[
\frac{R\Delta}{\pi(V)} + \left( 1 - \frac{R}{\pi(V)} \right)m(V) = m(V) + \frac{R}{\pi(V)}\left( \Delta - m(V) \right).
\tag{15a}\label{eq:15a}
\]

Thus equation~\eqref{eq:15a} yields the same augmented pseudo-event map used by
OVAC. The point estimator belongs to the established augmented Cox
family, most directly \citet{liu2010}; related augmented
inverse-probability ideas for incomplete Cox data also appear in Wang
and Chen (2001). OVAC does not claim a different augmented estimating
equation. Its contribution is to analyze that score in a cross-fitted
orthogonal framework, expose the exact product-form nuisance drift, and
derive the empirical risk-set linearization needed to connect the score
expansion to the observed-data efficient influence function. These
distinctions change the admissible nuisance-estimation strategy and the
implementation of analytic inference without obscuring the
estimator's algebraic ancestry.

\section{Identification, double robustness, and
orthogonality}

\textbf{Lemma 1 (Conditional augmentation identity).} For any integrable
scalar or vector \(H\) whose gold-standard value is observed only when
\(R = 1\), let \(m_{0}(V) = E(H \mid V)\) and define
\(\mathcal{A}(H;m,\pi) = m(V) + R\left( H - m(V) \right)/\pi(V)\). Under
\(R\bot H \mid V\),
\(E\left( \mathcal{A}(H;m,\pi) \mid V \right) = m(V) + \pi_{0}(V)\left( m_{0}(V) - m(V) \right)/\pi(V)\).

Lemma 1 is the generic missing-data identity underlying OVAC. Applying
it with \(H = \Delta\) gives a particularly sharp expression because
\(\Delta\) is binary and the Cox risk-set contrast is fully observed.

\textbf{Proposition 1 (Exact bias factorization).} For any \(\beta\) and
nuisance functions \(p\) and \(\pi\) with \(\pi\) bounded away from
zero,

\[
\Psi(\beta;p,\pi) - \Psi_{F}(\beta) = E\left\lbrack c_{\beta}(U,X)\left( 1 - \frac{\pi_{0}(V)}{\pi(V)} \right)\left( p(V) - p_{0}(V) \right) \right\rbrack.
\tag{16}\label{eq:16}
\]

Equation~\eqref{eq:16} shows that nuisance misspecification enters through the
product of validation-probability error and event-bridge error.
Consequently, first-order nuisance drift vanishes at the intersection
model and the population score remains unbiased whenever either nuisance
component is correct.

\textbf{Proposition 2 (Double-robust identification).} Assume the
full-data Cox model identifies \(\beta_{0}\). Then
\(\Psi(\beta;p,\pi) = \Psi_{F}(\beta)\) for every \(\beta\) if either
\(p = p_{0}\) almost surely or \(\pi = \pi_{0}\) almost surely.
Consequently \(\beta_{0}\) is identified by the OVAC score under the
union model \(\mathcal{M}_{p} \cup \mathcal{M}_{\pi}\).

\textbf{Lemma 2 (Neyman orthogonality).} Let \(\eta = (p,\pi)\) and
\(\eta_{0} = \left( p_{0},\pi_{0} \right)\). For square-integrable
perturbations \(h_{p}\) and \(h_{\pi}\), the Gateaux derivative of
\(\eta \mapsto \Psi\left( \beta_{0};\eta \right)\) at \(\eta_{0}\) in
either nuisance direction is zero. Equivalently, define
\(f_{p}(r) = \Psi\left( \beta_{0};p_{0} + rh_{p},\pi_{0} \right)\) and
\(f_{\pi}(r) = \Psi\left( \beta_{0};p_{0},\pi_{0} + rh_{\pi} \right)\).
Then \(f_{p}'(0) = 0\) and \(f_{\pi}'(0) = 0\).

The exact factorization yields the rate statement immediately. With
norms defined under the phase-I distribution and \(\widehat{\pi}\)
uniformly bounded below by \(\varepsilon/2\) with high probability,

\[
\left. \parallel\Psi\left( \beta_{0};\widehat{p},\widehat{\pi} \right) \right.\parallel \leq C\,\left. \parallel\widehat{p} - p_{0} \right.\parallel_{2}\left. \parallel\widehat{\pi} - \pi_{0} \right.\parallel_{2}.
\tag{17}\label{eq:17}
\]

for a finite constant \(C\) depending on the covariate envelope and
positivity bound. Therefore a product rate
\(o_{p}\left( n^{- 1/2} \right)\) is sufficient for nuisance bias to be
negligible at root-\(n\) scale. The common symmetric sufficient
condition
\(\parallel \widehat{p} - p_{0} \parallel_{2} = o_{p}\left( n^{- 1/4} \right)\)
and
\(\parallel \widehat{\pi} - \pi_{0} \parallel_{2} = o_{p}\left( n^{- 1/4} \right)\)
is only one possibility; one nuisance may converge more slowly if the
other converges faster. If \(\pi_{0}\) is known exactly by design, the
population remainder is identically zero for any \(p\).

\subsection{Why the result is specific to accurately observed risk
sets}

The clean product remainder depends on \(c_{\beta}(U,X)\) being
observable for everyone. If \(U\) is mismeasured, then \(c_{\beta}\)
itself depends on latent risk-set membership and cannot simply be pulled
outside the conditional expectation in Lemma 1. One would then need to
augment the at-risk process as well as the event increment, returning to
the more general measurement-error and raking literature. This
distinction is central to the scope of the method and prevents the
theoretical claims from being overextended.

\section{Large-sample theory}

\subsection{Regularity conditions}

The asymptotic results combine standard Cox-model conditions with
conditions for cross-fitted nuisance estimation \citep{andersen1982,chernozhukov2018,fleming1991}. A convenient
sufficient set is stated here; Appendix B records a more technical
version.

\begin{itemize}
\item
  (A1) The observations are i.i.d. copies of \(O = (V,R,R\Delta)\), with
  finite study horizon \(\tau\).
\item
  (A2) The Cox model in (1) holds, and censoring is conditionally
  independent of \(T^{0}\) given \(X\); the at-risk probability is
  bounded away from zero on \(\lbrack 0,\tau\rbrack\) on the relevant
  covariate support.
\item
  (A3) X is bounded almost surely, or has exponential moments sufficient
  for the standard Cox empirical-process arguments.
\item
  (A4) \(s^{(0)}(t,\beta)\) is uniformly bounded away from zero near
  \(\beta_{0}\); \(s^{(k)}\) are continuously differentiable in
  \(\beta\).
\item
  (A5) The information matrix
  \(I_{0} = E\left( \Delta v\left( U,\beta_{0} \right) \right)\) is
  positive definite.
\item
  (A6) Validation is missing at random given \(V\) and
  \(\pi_{0}(V) \geq \varepsilon > 0\).
\item
  (A7) The cross-fitted nuisance estimators satisfy
  \(0\leq\widehat p\leq 1\) and
  \(\widehat{\pi}\geq\varepsilon/2\) with probability tending to one,
  \(\parallel \widehat{p} - p_{0} \parallel_{2} = o_{p}(1)\),
  \(\parallel \widehat{\pi} - \pi_{0} \parallel_{2} = o_{p}(1)\), and
  \(\parallel \widehat{p} - p_{0} \parallel_{2}
   \parallel \widehat{\pi} - \pi_{0} \parallel_{2}
   = o_{p}\left( n^{- 1/2} \right)\).
\item
  (A8) Each held-out fold contains a nonvanishing fraction of the sample
  and the number of folds K is fixed.
\end{itemize}

\textbf{Lemma 3 (Derivative of the population score).} Under (A1)-(A5),
\(\Psi_{F}\) is differentiable at \(\beta_{0}\) and

\[
- \left. \ \frac{\partial\Psi_{F}(\beta)}{\partial\beta^{\top}} \right|_{\beta = \beta_{0}} = I_{0} = E\left( \Delta v\left( U,\beta_{0} \right) \right) = \int_{0}^{\tau}v\left( t,\beta_{0} \right)s^{(0)}\left( t,\beta_{0} \right)\, d\Lambda_{0}(t).
\tag{18}\label{eq:18}
\]

The second equality is the usual Cox information identity. Positive
definiteness supplies local identification and a stable first-order
expansion around the target.

\textbf{Theorem 1 (Consistency under the union model).} Suppose
(A1)-(A6) hold, \(\beta_{0}\) is the unique zero of \(\Psi_{F}\) on a
compact neighborhood \(B\), and the empirical risk-set processes
converge uniformly on \(B\). Assume the cross-fitted working nuisances
satisfy \(0\leq\widehat p\leq1\) and
\(\widehat\pi\geq\varepsilon/2\) with probability tending to one. If
either \(\parallel\widehat p-p_{0}\parallel_{2}=o_{p}(1)\) or
\(\parallel\widehat\pi-\pi_{0}\parallel_{2}=o_{p}(1)\), while the other
working nuisance remains bounded as just stated, then any approximate
root \(\widehat{\beta}\) satisfying
\(\parallel U_{n}\left( \widehat{\beta} \right) \parallel = o_{p}(1)\)
converges in probability to \(\beta_{0}\).

Theorem 1 concerns consistency under the union model. If only one
nuisance component is consistently estimated, the first-order
distribution can depend on estimation of the other component.
Accordingly, the efficient influence-function variance developed in
Section 6 is an intersection-model result: both \(p\) and \(\pi\) must be
consistently estimated, except that \(\pi\) may be known by design. Under
one-sided robustness, inference should instead use a subject-level
bootstrap that refits the nuisance functions and resolves the OVAC
score.

\textbf{Theorem 2 (Cross-fitted asymptotic linearity).} Under (A1)-(A8),
suppose \(\widehat\beta\) is an approximate solution satisfying
\(\parallel U_n(\widehat\beta)\parallel=o_p(n^{-1/2})\). Then the OVAC
estimator admits the expansion

\[
\sqrt{n}\left( \widehat{\beta} - \beta_{0} \right) = I_{0}^{- 1}\frac{1}{\sqrt{n}}\sum_{i = 1}^{n}\xi_{0}\left( O_{i} \right) + o_{p}(1).
\tag{19}\label{eq:19}
\]

where \(\xi_{0}(O)=D_{0}c_{0}(U,X)-G_{0}(U,X)\), with \(c_{0}(t,X)=X-\bar{X}(t,\beta_{0})\) and \(G_{0}\) defined in
Section 6. The term \(G_{0}\) is the first-order contribution from estimating
the empirical Cox risk-set mean. It remains in the score expansion after
\(\Delta\) is replaced by an augmented pseudo-event. Consequently, \(\widehat{\beta}\) is root-\(n\)
consistent and asymptotically normal.

\textbf{Risk-set linearization identity.} At the intersection model, the pseudo-event score evaluated at the true nuisance
functions satisfies

\[\sqrt{n}\, U_{n}\left( \beta_{0},\eta_{0} \right) = \frac{1}{\sqrt{n}}\sum_{i = 1}^{n}\left\lbrack D_{0i}c_{0}\left( U_{i},X_{i} \right) - G_{0}\left( U_{i},X_{i} \right) \right\rbrack + o_{p}(1).\]

This identity is the link between the empirical pseudo-event score and
the efficient influence function. A direct substitution of the
pseudo-event for a martingale jump would miss the risk-set contribution.
Appendix A derives this identity explicitly by expanding the empirical risk-set mean,
reducing the resulting second-order term to its first-order Hoeffding projection,
and identifying that projection with the $G_0$ contribution.

\textbf{Corollary 1 (Asymptotic normality).} Under the conditions of
Theorem 2,

\[
\sqrt{n}\left( \widehat{\beta} - \beta_{0} \right)\  \Rightarrow \ N_{p}\left( 0,V_{eff} \right),\quad\quad V_{eff} = E\left( \phi_{eff}(O)\phi_{eff}(O)^{\top} \right).
\tag{20}\label{eq:20}
\]

The proof follows the standard orthogonal Z-estimation decomposition: a
true-nuisance empirical-process term, a cross-fitted empirical nuisance term,
and the product-form population remainder, followed by linearization in
\(\beta\) using Lemma 3. Appendix A gives the details.

\subsection{Rates when validation probabilities are
known}

Designed validation studies often know \(\pi_{0}\) exactly. In that case
Proposition 1 implies
\(\Psi\left( \beta;p,\pi_{0} \right) = \Psi_{F}(\beta)\) for every
working bridge \(p\). Consistency therefore does not require a correct
event model. Efficiency, however, still depends on \(p\) approaching
\(p_{0}\) because the augmentation term is an efficiency device as well
as a bias-correction device. With known \(\pi_{0}\) and cross-fitting, a
poor but bounded \(p\) can preserve consistency while sacrificing
variance; a high-quality bridge moves the estimator toward the
semiparametric bound. This separation of robustness and efficiency is
useful in practice because investigators can safely deploy flexible
classifiers without making their correctness a condition for point
identification when the sampling design is known.

\section{Semiparametric efficiency and variance
estimation}

\subsection{Full-data efficient influence
function}

Let \(c_{0}(t,X)=X-\bar{X}(t,\beta_{0})\) and define the full-data martingale

\[
M(t) = N(t) - \int_{0}^{t}Y(u)\exp\left( \beta_{0}^{\top}X \right)\, d\Lambda_{0}(u).
\tag{21}\label{eq:21}
\]

\textbf{Proposition 3 (Full-data Cox influence function).} Under the Cox model
and standard regularity conditions, the efficient influence function for
\(\beta_{0}\) in the full-data semiparametric model is the usual counting-process
Cox influence function \citep{andersen1982,fleming1991}:

\[
\phi_{F}(F) = I_{0}^{- 1}\int_{0}^{\tau}c_{0}(t,X)\, dM(t).
\tag{22}\label{eq:22}
\]

Because each subject has at most one event, this can be written

\[
\phi_{F}(F) = I_{0}^{- 1}\left\lbrack \Delta c_{0}(U,X) - G_{0}(U,X) \right\rbrack.
\tag{23}\label{eq:23}
\]

\[
G_{0}(U,X) = \exp\left( \beta_{0}^{\top}X \right)\int_{0}^{U}c_{0}(t,X)\, d\Lambda_{0}(t).
\tag{24}\label{eq:24}
\]

The second term depends only on \(U\), \(X\), \(\beta_{0}\), and \(\Lambda_{0}\) and is therefore
phase-I observable up to nuisance estimation. Only the event
contribution \(\Delta c_{0}(U,X)\) is hidden for nonvalidated subjects.

\subsection{Observed-data efficient influence
function}

\textbf{Lemma 4 (Coarsening-at-random projection).} Let the full data
be \(F=(V,H)\), where only \(H\) can be missing, and suppose
\(R\bot H\mid V\) with \(\Pr(R=1\mid V)=\pi_{0}(V)>0\). For a mean-zero
full-data canonical gradient \(\phi_{F}(F)\), the observed-data
canonical gradient is
\[
E\left( \phi_{F} \mid V \right)
+\frac{R}{\pi_{0}(V)}
\left\{\phi_{F}-E\left( \phi_{F} \mid V \right)\right\}.
\]

The projection in Lemma 4 is the standard coarsening-at-random construction
from semiparametric missing-data theory \citep{bickel1993,kennedy2016,robins1994,tsiatis2006}. In the current model,
\(E(\Delta\mid V)=p_{0}(V)\), so the projection has a closed form.

\textbf{Theorem 3 (Efficient influence function for internally validated
failure indicators).} The observed-data efficient influence function for
\(\beta_{0}\) is

\[
\phi_{eff}(O) = I_{0}^{- 1}\left\lbrack D_{0}c_{0}(U,X) - G_{0}(U,X) \right\rbrack.
\tag{25}\label{eq:25}
\]

\[
D_{0} = p_{0}(V) + \frac{R}{\pi_{0}(V)}\left( \Delta - p_{0}(V) \right).
\tag{26}\label{eq:26}
\]

Equation (25) is the central efficiency representation. The coarsening
projection replaces the hidden event contribution \(\Delta c_{0}(U,X)\) with its
augmented counterpart, but the compensator \(G_{0}\) is unchanged because it is
phase-I measurable. The same \(G_{0}\) term arises independently from
linearizing the empirical Cox risk-set mean in Theorem 2. This agreement
connects the cross-fitted score expansion to the observed-data efficient
influence function; \(G_{0}\) therefore belongs in both the asymptotic
expansion and an analytic variance estimator intended to represent that
score. Its first-order status concerns the influence representation
itself, not the sign or size of the net variance change obtained by
deleting it. Section 9.3 shows that \(G_{0}\) can be large as a component while
its variance contribution is nearly canceled by covariance with the
event-score component in a particular design.

\textbf{Theorem 4 (Efficiency attainment).} Under the conditions of
Theorem 2, suppose its asymptotic linear expansion is locally uniform
along regular parametric submodels through the true observed-data law.
Then the OVAC estimator is regular and asymptotically linear with
influence function \(\phi_{eff}\). Hence its asymptotic covariance
\(V_{eff}\) equals the semiparametric efficiency bound in the
observed-data model.

\subsection{Efficiency gain from
augmentation}

The role of the event bridge can be seen without solving the Cox model.
Consider the generic augmented influence class

\[
\phi_{h} = h(V) + \frac{R}{\pi_{0}(V)}\left( \phi_{F} - h(V) \right).
\tag{27}\label{eq:27}
\]

\textbf{Proposition 4 (Optimal augmentation within the
inverse-probability class).} Among square-integrable \(h(V)\),
\(Var\left( \phi_{h} \right)\) is minimized in Loewner order by
\(h_{0}(V) = E\left( \phi_{F} \mid V \right)\). For any scalar contrast
\(a\), the excess variance of the unaugmented Horvitz-Thompson influence
\(R\phi_{F}/\pi_{0}\) over the efficient influence is

\[
E\left\lbrack \left( \frac{1}{\pi_{0}(V)} - 1 \right)\left( a^{\top}h_{0}(V) \right)^{2} \right\rbrack \geq 0.
\tag{28}\label{eq:28}
\]

Thus, when the conditional mean is correctly specified, augmentation
cannot increase asymptotic variance within this inverse-probability
class. The gain is largest when phase-I information strongly predicts
the complete-data influence and the validation fraction is limited.

\subsection{Practical variance
estimator}

At the efficient intersection model, the analytic variance estimator
should mirror the empirical score linearization. Let
\({\widehat{c}}_{i}(t) = X_{i} - \bar{X}_{n}\left( t,\widehat{\beta} \right)\),
and let
\(\widehat{I} = n^{- 1}\sum_{i}^{}{\widetilde{D}}_{i}\widehat{v}\left( U_{i},\widehat{\beta} \right)\).
Define the signed pseudo-event linearization measure

\[
d{\widehat{A}}_{D}(t) = \frac{\sum_{i:U_{i} = t}^{}{\widetilde{D}}_{i}}{\sum_{j = 1}^{n}Y_{j}(t)\exp\left( {\widehat{\beta}}^{\top}X_{j} \right)}.
\tag{29}\label{eq:29}
\]

The increments in (29) need not be nonnegative because \(\widetilde{D}\)
is an augmented pseudo-event; \({\widehat{A}}_{D}\) is therefore a
score-linearization measure, not a finite-sample baseline-hazard
estimator. Under the intersection model it converges to \(\Lambda_{0}\)
because \(E\left( D_{0}f(U,X) \right) = E\left( \Delta f(U,X) \right)\)
for phase-I-measurable \(f\). Define

\[
{\widehat{G}}_{i} = \exp\left( {\widehat{\beta}}^{\top}X_{i} \right)\int_{0}^{U_{i}}{\widehat{c}}_{i}(t)\, d{\widehat{A}}_{D}(t).
\tag{30}\label{eq:30}
\]

\[
{\widehat{\phi}}_{i} = {\widehat{I}}^{- 1}\left( {\widetilde{D}}_{i}{\widehat{c}}_{i}\left( U_{i} \right) - {\widehat{G}}_{i} \right).
\tag{31}\label{eq:31}
\]

\[
\widehat{Var}\left( \widehat{\beta} \right) = \frac{1}{n^{2}}\sum_{i = 1}^{n}\left( {\widehat{\phi}}_{i} - \bar{\widehat{\phi}} \right)\left( {\widehat{\phi}}_{i} - \bar{\widehat{\phi}} \right)^{\top}.
\tag{32}\label{eq:32}
\]

\textbf{Proposition 5 (Consistency of the analytic variance estimator).}
Under (A1)--(A8), the bounded-covariate version of (A3) (or moment and
stochastic-equicontinuity conditions yielding the same uniform laws),
the bounded-variation condition on \(t\mapsto\bar{X}_0(t)\) stated in
Appendix B, and the intersection model,
\[
 n\,\widehat{Var}(\widehat\beta)\ \overset{p}{\longrightarrow}\ 
 V_{eff}.
\]
Thus the standard errors obtained from (32) are first-order consistent
for the sampling standard errors of \(\widehat\beta\).

A bridge-stabilized alternative replaces D in (29) by p. When p is
consistent, this estimates the same population \(G_{0}\) and is useful as an
intersection-model diagnostic, but it does not directly linearize the
empirical pseudo-event score. The accompanying implementation therefore
uses the direct pseudo-event score-linearization measure by default.
Under one-sided double robustness with an estimated active nuisance
model, additional nuisance influence terms may enter; unless those terms
are derived explicitly, a subject-level bootstrap that refits all
nuisances and resolves the OVAC root is the recommended inferential
procedure.

\section{Efficient validation
design}

Validation is often the expensive component of an endpoint-error study.
The influence-function representation makes it possible to design phase
II directly for the regression contrast of interest. Let
\(a \in \mathbb{R}^{p}\) be fixed and consider
\(\theta = a^{\top}\beta_{0}\). Define

\[
\sigma_{a}^{2}(V) = Var\left( a^{\top}\phi_{F}(F) \mid V \right).
\tag{33}\label{eq:33}
\]

Under the present model, only \(\Delta\) is missing conditional on V, so (23)
gives

\[
\sigma_{a}^{2}(V) = \left( a^{\top}I_{0}^{- 1}c_{0}(U,X) \right)^{2}p_{0}(V)\left( 1 - p_{0}(V) \right).
\tag{34}\label{eq:34}
\]

Suppose the investigator can choose a Bernoulli validation probability
\(\pi(V)\) subject to the expected budget \(E\left( \pi(V) \right) = q\)
and the design positivity constraint \(\pi_{\min} \leq \pi(V) \leq 1\).
The efficient variance for \(\theta\) decomposes as

\[
Var\left( a^{\top}\phi_{eff} \right) = Var\left\lbrack E\left( a^{\top}\phi_{F} \mid V \right) \right\rbrack + E\left( \frac{\sigma_{a}^{2}(V)}{\pi(V)} \right).
\tag{35}\label{eq:35}
\]

\textbf{Theorem 5 (Optimal validation allocation).} Fix a design lower
bound \(\pi_{\min} \in (0,1)\) and an interior budget
\(q\in(\pi_{\min},1)\). Suppose
\(E\{\sigma_a(V)\}<\infty\). Among measurable validation probabilities
satisfying \(E\{\pi(V)\}=q\) and
\(\pi_{\min}\leq\pi(V)\leq1\), every minimizer agrees on
\(\{\sigma_a(V)>0\}\) with

\[
\pi_{opt}(V) = \min\left[1,\max\left\{\pi_{\min},
\kappa\sigma_{a}(V)\right\}\right],
\tag{36}\label{eq:36}
\]

for a multiplier \(\kappa>0\) chosen to satisfy the budget whenever
\(\Pr\{\sigma_a(V)>0\}=1\). If \(\sigma_a(V)=0\) on a set of positive
probability, any residual budget after applying the clipped rule on
\(\{\sigma_a>0\}\) may be assigned arbitrarily on that zero-variance set
without changing the objective. If neither bound is active and
\(\Pr\{\sigma_a(V)>0\}=1\), then
\(\pi_{opt}(V)=q\sigma_a(V)/E\{\sigma_a(V)\}\). The endpoint budgets
\(q=\pi_{\min}\) and \(q=1\) are attained by the constant rules
\(\pi\equiv\pi_{\min}\) and \(\pi\equiv1\), respectively.

\textbf{Corollary 2 (Interpretation for a single Cox coefficient).} For
\(\theta = \beta_{0j}\), optimal validation preferentially samples
records for which \(\left| e_{j}^{\top}I_{0}^{- 1}c_{0}(U,X) \right|\)
is large and \(p_{0}(V)\) is near \(1/2\). Thus the design targets both
statistical leverage and classification ambiguity.

This design principle differs from simply oversampling apparent cases.
If \(\widetilde{\Delta} = 1\) almost perfectly predicts the true event,
then \(p_{0}\left( 1 - p_{0} \right)\) is small and further validation
of obvious apparent cases contributes little information about
classification. Records with uncertain classification can be more
valuable, especially when their covariate pattern exerts large influence
on the target coefficient. Conversely, a highly ambiguous record with
nearly zero Cox contrast contributes little to the coefficient and need
not be prioritized. Equation~\eqref{eq:36} formalizes the tradeoff.

\subsection{Adaptive multi-wave
implementation}

The optimal rule depends on unknown \(\beta_{0}\), \(I_{0}\),
\(\bar{X}\), and \(p_{0}\), so a fully optimal design is unavailable before
any validation occurs. A practical approach is multi-wave sampling. An initial
pilot wave can be selected by simple or stratified random sampling, after which
preliminary estimates of \(p_{0}\) and the influence components are used to
update \(\pi\) for the next wave. This
strategy follows the broader influence-function and adaptive
validation-design literature \citep{han2021,shepherd2023,tao2021}, while the present model yields the unusually simple
target in (34). To preserve valid inference, the cumulative inclusion
probability for each record must be tracked across waves, and subsequent
analysis should condition on the known or consistently estimated final
inclusion mechanism.

\section{Interpretation, causal use, and
extensions}

\subsection{Association parameter versus causal hazard
ratio}

The statistical theory targets a Cox regression coefficient, not
automatically a causal effect. If one component of X is an exposure or
treatment A and the remaining components are baseline confounders L,
causal interpretation requires additional assumptions: consistency,
conditional exchangeability given L, treatment positivity, an
appropriate independent-censoring condition, correct handling of
post-baseline treatment changes, and a structural proportional-hazards
interpretation suited to the scientific question. Measurement error can
also be represented as an additional missing-data problem in a
potential-outcomes framework \citep{edwards2015}. Even under causal
identification assumptions, hazard ratios require careful interpretation
and should not be read as risk ratios \citep{hernan2010}. OVAC addresses
endpoint classification; it does not correct unmeasured confounding or
rescue an unsuitable causal estimand.

\subsection{Time-dependent
covariates}

The pseudo-event idea extends to predictable external covariate
processes \(X(t)\) provided the process is accurately observed for all
subjects and the risk sets remain known. Replace \(c_{\beta}(U,X)\) by
the event-time contrast \(X(U) - \bar{X}(U,\beta)\), where
\(s^{(k)}(t,\beta) = E\left\lbrack Y(t)\exp\left( \beta^{\top}X(t) \right)X(t)^{\otimes k} \right\rbrack\).
The augmentation identity is unchanged because it acts on \(\Delta\).
Establishing the full empirical-process theory requires the standard
predictable-process conditions for the extended Cox model, but no new
measurement-error argument is needed.

\subsection{Misclassified causes of
failure}

For competing risks with accurately observed all-cause event times but an
error-prone cause label, the same structure can be applied cause by cause. For
cause \(k\), the hidden indicator is
\(\Delta_{k}=I(\text{event cause}=k)\), while \(Y(t)\) remains observed. A
cause-specific OVAC score can therefore replace \(\Delta_{k}\) by an augmented
pseudo-cause. However, causes are multinomial rather than binary and the
joint covariance across cause-specific estimators must account for the
simplex constraint. Existing Cox work on misclassified causes of failure
provides the primary comparison \citep{vanrompaye2012}, so this
extension is not claimed as a new result here.

\subsection{Why mismeasured event times require new
theory}

If \(U\) itself is error-prone, then \(Y(t)\) is latent and the risk-set mean
\({\bar{X}}_{n}(t,\beta)\) is no longer a phase-I observable. Correcting only
the event indicator would generally be inconsistent. One would need to model
or calibrate the latent risk-set process, use a likelihood or imputation
strategy for event-time error, or construct an augmented counting-process
estimator that jointly recovers \(Y\) and \(N\). Generalized raking, regression calibration,
SIMEX, quasi-likelihood, and internal-validation methods for broader
survival measurement error provide natural starting points \citep{boe2021,giganti2020,oh2018,oh2021a,oh2021b,zucker2008,zucker2019}. Related recent work also treats
risk and survival functions under outcome measurement error (Edwards et
al., 2026). The present paper deliberately avoids claiming that OVAC
solves that broader problem.

\section{Monte Carlo experiments}

\subsection{Data-generating process}

We used a common data-generating mechanism with $n=2{,}000$ subjects across three
Monte Carlo exercises: a 250-replication method-comparison experiment, a
1,000-replication R variance-calibration experiment, and a 200-replication
component diagnostic for the risk-set term. The 250-replication experiment
examined bias from treating the error-prone failure indicator as truth,
recovery of the full-data Cox target, and efficiency relative to
inverse-probability weighting; the larger R experiment was used for the
primary analytic-variance calibration. Covariates $X_1$ and $Z$ were generated
independently from standard normal distributions.
Conditional event times were exponential with rate

\[
\lambda_{T}\left( X_{1},Z \right) = 0.08\exp\left( 0.5X_{1} - 0.3Z \right),
\tag{37}\label{eq:37}
\]

so that the population Cox coefficient was
\(\beta_{0} = (0.5, - 0.3)^{\top}\). Censoring times were exponential
with rate

\[
\lambda_{C}(Z) = 0.06\exp(0.25Z).
\tag{38}\label{eq:38}
\]

The error-prone event indicator had sensitivity 0.82 and specificity
0.90. Validation sampling depended on phase-I information according to

\[
logit\left( \pi_{0}(V) \right) = - 1.55 + 1.10\widetilde{\Delta} + 0.35X_{1},
\tag{39}\label{eq:39}
\]

which yielded a validation fraction of roughly 0.29. The event bridge
used \(\widetilde{\Delta}\), \(X_{1}\), and \(Z\), and the validation
model used \(\widetilde{\Delta}\) and \(X_{1}\). Five-fold cross-fitting
was used for the nuisance regressions. Within each replication of the 250-replication method-comparison experiment,
each generated survival sample was analyzed with four regression models: the
Full-data Cox regression model (FDC), using the true failure indicator for every
subject; the Naive Cox regression model, treating \(\widetilde{\Delta}\) as
correct; the inverse-probability-weighted Cox regression model (IPW), using the
validation sample and validation probabilities; and the OVAC regression model.
FDC is an infeasible full-information benchmark used to isolate the cost of
incomplete endpoint validation, not a candidate analysis when validation is
incomplete.

The method-comparison experiment in Tables~\ref{tab:mc-beta1} and~\ref{tab:mc-beta2} uses 250 replications
under the correctly specified intersection model and evaluates point
recovery and efficiency relative to FDC, Naive, and IPW analyses.
Because the principal inferential claim concerns the risk-set-linearized
variance, Section 9.3 promotes a separate 1,000-replication R run under
the same data-generating mechanism and five-fold cross-fitting
specification to the primary variance-calibration experiment. A
200-replication component diagnostic then decomposes the event-score and
risk-set contributions, and an independently coded Python implementation
provides a cross-language reproducibility check. Double-robust
identification does not require both nuisance models to be correct, but
under one-sided correctness the first-order distribution can contain
nuisance-estimation contributions not represented by the
intersection-model formula. The recommended procedure in that setting is
a subject-level bootstrap that repeats nuisance fitting, cross-fitting,
and solution of the OVAC score.

\subsection{Point-estimation
performance}

Tables~\ref{tab:mc-beta1} and~\ref{tab:mc-beta2} report performance separately for the two regression
coefficients. Bias is measured against the population parameter, while
the paired difference from FDC compares methods on the same generated
survival sample. The two summaries answer different questions: population
bias evaluates recovery of the statistical target, whereas the paired FDC
discrepancy isolates the incremental effect of incomplete endpoint validation
and misclassification within each replicate.

\begin{table}[htbp]
\centering
\caption{Monte Carlo performance for $\beta_1=0.5$ ($B=250$, $n=2{,}000$).}
\label{tab:mc-beta1}
\small
\begin{tabular}{lrrrrrr}
\toprule
Method & Mean & Bias vs. truth & Empirical SD & Mean SE & Coverage & RMSE \\
\midrule
FDC    & 0.49955 & -0.00045 & 0.03059 & 0.03162 & 0.960 & 0.03053 \\
Naive  & 0.45780 & -0.04220 & 0.03375 & 0.03339 & 0.744 & 0.05400 \\
IPW    & 0.50322 &  0.00322 & 0.05739 & 0.05784 & 0.952 & 0.05736 \\
OVAC   & 0.49967 & -0.00033 & 0.04488 & 0.04449 & 0.960 & 0.04479 \\
\bottomrule
\end{tabular}
\end{table}

For \(\beta_{1}\), the Naive estimator was attenuated by approximately
0.042 on the log-hazard scale and its nominal 95\% interval covered only
74.4\% of replications. IPW removed most of the bias but paid a
substantial variance penalty. The OVAC mean was 0.49967, while the
paired FDC mean was 0.49955; the difference in means was 0.00013. The
paired RMSE of OVAC relative to FDC was 0.03313.

\begin{table}[htbp]
\centering
\caption{Monte Carlo performance for $\beta_2=-0.3$ ($B=250$, $n=2{,}000$).}
\label{tab:mc-beta2}
\small
\begin{tabular}{lrrrrrr}
\toprule
Method & Mean & Bias vs. truth & Empirical SD & Mean SE & Coverage & RMSE \\
\midrule
FDC    & -0.30038 & -0.00038 & 0.03059 & 0.03082 & 0.944 & 0.03053 \\
Naive  & -0.25151 &  0.04849 & 0.03110 & 0.03255 & 0.684 & 0.05757 \\
IPW    & -0.30225 & -0.00225 & 0.05254 & 0.05415 & 0.956 & 0.05249 \\
OVAC   & -0.30426 & -0.00426 & 0.04148 & 0.04123 & 0.956 & 0.04161 \\
\bottomrule
\end{tabular}
\end{table}

For \(\beta_{2}\), Naive Cox regression shifted the coefficient toward
zero by approximately 0.0485 and produced only 68.4\% coverage. OVAC
retained a small finite-sample bias of -0.00426 relative to the
population target and differed from the paired FDC mean by -0.00388.
Its paired RMSE relative to FDC was 0.02866. These results illustrate why
both population-target bias and paired FDC discrepancy are useful summaries:
the former evaluates the statistical target, whereas the latter isolates the
incremental effect of incomplete endpoint validation within the same generated
survival samples.

\subsection{Variance calibration}

The analytic variance in Section 6.4 retains the first-order risk-set
contribution \(G_{0}\) implied by Theorem 2. Table~\ref{tab:variance-calibration} reports the primary
variance calibration from 1,000 R replications under the Section 9
data-generating mechanism. The principal diagnostics are the ratio of
mean analytic standard error to empirical Monte Carlo standard deviation
and nominal 95\% coverage. The earlier 250-replication method-comparison
run is retained in Tables~\ref{tab:mc-beta1}, \ref{tab:mc-beta2}, and~\ref{tab:efficiency} because it contains all four
estimators; it is no longer the headline variance-calibration result.

\begin{table}[htbp]
\centering
\caption{Primary 1,000-replication R calibration of the full risk-set-linearized OVAC variance.}
\label{tab:variance-calibration}
\small
\begin{tabular}{lrrrrr}
\toprule
Term & Mean $\widehat\beta$ & MC SD & Mean full SE & SE/SD & 95\% coverage \\
\midrule
$\beta_1$ ($x_1$) & 0.5015 & 0.0446 & 0.04443 & 0.997 & 0.955 \\
$\beta_2$ ($z$)   & -0.3028 & 0.0416 & 0.04122 & 0.992 & 0.948 \\
\bottomrule
\end{tabular}
\end{table}

The full score-linearization standard errors closely tracked Monte Carlo
variation: SE/SD was 0.997 for \(\beta_{1}\) and 0.992 for \(\beta_{2}\), with coverage of
0.955 and 0.948. With 1,000 replications, the Monte Carlo standard error
of a coverage estimate near 0.95 is approximately 0.007. The \(\beta_{1}\) SE/SD
ratio differs from the 0.991 value in the earlier 250-replication run
because the empirical Monte Carlo SD changed from 0.04488 to
approximately 0.0446; at $B=250$ the relative Monte Carlo uncertainty of
an SD estimate is about 4.5\%, compared with about 2.2\% at $B=1{,}000$.
The two calibration runs are therefore statistically compatible, but the
larger run is the more precise basis for the inferential claim.

The ablation must be interpreted through the component structure rather than through the net SE change alone. Let \(\phi_D\) denote the coefficient-specific influence contribution obtained from \(\widehat I^{-1}\widetilde D_i\widehat c_i(U_i)\), and let \(\phi_G\) denote the contribution obtained from \(\widehat I^{-1}\widehat G_i\), so that the full contribution is \(\phi_D-\phi_G\). By construction, \(\sum_i\widehat G_i=0\) because the empirical risk-set identity \(\widehat S_1(t)-\widehat S_0(t)\bar{X}(t)=0\) holds pointwise. Deleting \(G_0\) therefore leaves score centering and the OVAC point estimate unchanged; it can affect only second moments. For each coefficient,
\[
\operatorname{Var}(\phi_D-\phi_G)=\operatorname{Var}(\phi_D)+\operatorname{Var}(\phi_G)-2\operatorname{Cov}(\phi_D,\phi_G).
\]
Table~\ref{tab:g0-components} reports a 200-replication diagnostic of these components. That table is a component-magnitude diagnostic, not the source of the paired no-$G_0$ ablation percentages reported below. The 0.173\% and 0.223\% R ablation changes come from the separate 1,000-replication R calibration, in which the full and no-$G_0$ variance estimates were compared within the same replication. Consequently, the rounded Full-SE and $\widetilde D\widehat c$-only entries in Table~\ref{tab:g0-components} should not be used to reconstruct those 1,000-replication percentages.

\begin{table}[htbp]
\centering
\caption{Component decomposition of the $G_0$ influence contribution in a 200-replication diagnostic.}
\label{tab:g0-components}
\small
\begin{tabular}{lrrrrr}
\toprule
Term & Full SE & $\widetilde D\widehat c$-only SE & $\widehat G$-only SE & $\widehat G$/full & $\mathrm{Corr}(\phi_D,\phi_G)$ \\
\midrule
$\beta_1$ ($x_1$) & 0.0442 & 0.0442 & 0.0344 & 77.8\% & 0.395 \\
$\beta_2$ ($z$)   & 0.0415 & 0.0417 & 0.0337 & 81.1\% & 0.412 \\
\bottomrule
\end{tabular}
\end{table}

The decomposition shows that $G_0$ is not numerically small. Processed through $\widehat I^{-1}$ on its own, the $G_0$ component has an SE equal to 77.8\% of the full SE for $\beta_1$ and 81.1\% for $\beta_2$. The coefficient-specific correlations between $\phi_D$ and $\phi_G$ are 0.395 and 0.412, so the cross-covariance term nearly offsets $\operatorname{Var}(\phi_G)$. This is why deleting $G_0$ has almost no net effect in this DGP. In the 1,000-replication R run, the no-$G_0$ variance increased mean SE by 0.173\% for $\beta_1$ and 0.223\% for $\beta_2$; an independent Python implementation gave increases of 0.157\% and 0.236\%. The slight increase rather than decrease occurs because $2\operatorname{Cov}(\phi_D,\phi_G)$ marginally exceeds $\operatorname{Var}(\phi_G)$ in this calibration regime. There is no theoretical guarantee that this cancellation, or even its sign, persists under different censoring, validation, event-frequency, or covariate regimes. The two implementations also agree closely on the full analytic quantities despite independent random-number streams and different GLM routines: mean full SEs were 0.04443 versus 0.04456 for $\beta_1$ and 0.04122 versus 0.04125 for $\beta_2$ in R and Python, respectively. Their full-variance coverage estimates (0.955/0.948 in R and 0.943/0.946 in Python) differ by amounts compatible with $B=1{,}000$ Monte Carlo noise; the visible $\beta_1$ SE/SD difference is driven by empirical SD estimates of about 0.0446 versus 0.0459, a gap within roughly one combined Monte Carlo SD-estimation error. Thus the ablation is evidence of covariance cancellation, not evidence that $G_0$ is ignorable.

\subsection{Efficiency relative to inverse-probability
weighting}

Augmentation substantially reduced sampling variability relative to IPW
in the 250-replication method-comparison experiment. For \(\beta_{1}\), empirical
variance fell by 38.8\% and RMSE fell by 21.9\%. For \(\beta_{2}\), empirical
variance fell by 37.7\% and RMSE fell by 20.7\%.

\begin{table}[htbp]
\centering
\caption{Efficiency of OVAC relative to inverse-probability weighting in the 250-replication method-comparison design.}
\label{tab:efficiency}
\small
\begin{tabular}{lrrr}
\toprule
Term & SD (IPW / OVAC) & RMSE (IPW / OVAC) & Reduction (variance / RMSE) \\
\midrule
$\beta_1$ ($x_1$) & 0.05739 / 0.04488 & 0.05736 / 0.04479 & 38.8\% / 21.9\% \\
$\beta_2$ ($z$)   & 0.05254 / 0.04148 & 0.05249 / 0.04161 & 37.7\% / 20.7\% \\
\bottomrule
\end{tabular}
\end{table}

The efficiency gain is expected when the phase-I variables predict the
true event indicator: IPW uses the validation sample primarily through
inverse sampling weights, whereas OVAC uses the event bridge to recover
information from nonvalidated records and uses validation residuals to
retain robustness. These numerical results support the
efficiency-over-IPW claim for the calibration design; they should not be
interpreted as a universal percentage gain, which depends on the
validation fraction, predictive strength of \(V\), positivity, event
frequency, and covariate distribution.

\subsection{Scope of the numerical
evidence}

The numerical evidence is intentionally focused rather than a
comprehensive stress test. Tables~\ref{tab:mc-beta1}, \ref{tab:mc-beta2}, and~\ref{tab:efficiency} use the original
250-replication method-comparison experiment to evaluate point recovery
and efficiency relative to alternative estimators. The primary
analytic-variance calibration is the separate 1,000-replication R
experiment in Table~\ref{tab:variance-calibration}, with independent Python verification. Table~\ref{tab:g0-components}
uses 200 replications to diagnose the mechanism behind the near-zero net
\(G_{0}\) ablation. That diagnostic is essential to interpretation: \(G_{0}\) is large
as an individual first-order component, but in this DGP $\operatorname{Var}(\phi_G)$ is
almost canceled by $2\operatorname{Cov}(\phi_D,\phi_G)$.  The cancellation should not be
extrapolated to other validation fractions, censoring regimes, event
frequencies, or covariate distributions. Semiparametric efficiency
follows from the influence-function argument under the stated model, not
from simulation, and the reported analytic coverage should not be
extrapolated to one-sided nuisance misspecification. Double robustness
is established theoretically by Proposition 2 and Theorem 1.
Misspecification, rare-event, validation-fraction, censoring, and
sample-size experiments remain natural extensions. The numerical claims
in Tables 1-5 should therefore be read only for the stated
data-generating mechanisms and simulation regimes.

\section{Practical implementation}

\subsection{Fitting the event
bridge}

The bridge model is fit only on validated records, but because
\(R\bot\Delta \mid V\) the conditional law of \(\Delta\) given \(V\) is
the same among validated and nonvalidated records. No
inverse-probability weight is required merely to estimate \(p_{0}(V)\)
as a conditional regression, although weights can be useful for
finite-sample stabilization or if the fitting algorithm targets a
marginal rather than conditional loss. The feature set should include
\(\widetilde{\Delta}\) and any phase-I variables that predict
discrepancies between \(\widetilde{\Delta}\) and \(\Delta\). Calibration
of predicted probabilities matters for efficiency because
\(\widehat{p}\) enters as a conditional mean rather than solely as a
rank score.

Flexible learners are useful when endpoint misclassification is
differential or nonlinear, and cross-fitting permits their use without
imposing a Donsker-class restriction. In moderate samples, however, a
well-specified low-dimensional logistic model may outperform a
high-variance learner. A practical strategy is to predefine a small
library of plausible learners, evaluate out-of-fold log loss or Brier
score within validated training data, and use an ensemble only when the
validation sample supports it. Tuning directly on the final Cox
coefficient should be avoided because it entangles nuisance selection
with target estimation.

\subsection{Validation
probabilities}

When validation is prospectively sampled with recorded inclusion
probabilities, those design probabilities should be used directly.
Estimating a second model for a known design adds unnecessary noise and
can weaken robustness. When \(\pi_{0}\) is unknown, logistic regression or
flexible classification of R on V can be used because R is observed for
the entire cohort. Positivity should be examined at the design stage
whenever possible. Extreme validation weights are not merely a numerical
inconvenience; they identify regions of phase-I space in which the
gold-standard event process is weakly informed by validated records.

\subsection{Newton-Raphson update}

With D fixed, the score derivative has the ordinary Cox form with
pseudo-event weights:

\[
J_{n}(\beta) = \frac{1}{n}\sum_{i = 1}^{n}{\widetilde{D}}_{i}V_{n}\left( U_{i},\beta \right).
\tag{40}\label{eq:40}
\]

where
\(V_{n}(t,\beta) = S_{n}^{(2)}(t,\beta)/S_{n}^{(0)}(t,\beta) - \left( S_{n}^{(1)}(t,\beta)/S_{n}^{(0)}(t,\beta) \right)^{\otimes 2}\).
The update is
\(\beta^{new} = \beta^{old} + J_{n}\left( \beta^{old} \right)^{- 1}U_{n}\left( \beta^{old} \right)\).
Because some \({\widetilde{D}}_{i}\) can be negative, \(J_{n}\) need not
be positive definite far from the solution in small samples. A damped
Newton step or a small ridge stabilization can be used computationally;
these devices should vanish asymptotically and should be reported if
materially active.

\subsection{Software architecture}

An R implementation can use the standard Cox risk-set machinery while
solving the pseudo-event score directly. This is preferable to treating
a conventional coxph call as a black box because augmented pseudo-events
can be negative and standard interfaces generally expect binary event
status. A production implementation should separate nuisance
cross-fitting, score solution, risk-set-linearized influence-function
inference, and validation-design calculations. Unit tests should recover
ordinary Cox regression when every subject is validated, recover IPW
when the bridge contribution is set to zero, and verify large-sample
invariance to the working bridge when validation probabilities are known
by design.

\section{Discussion}

This paper develops a cross-fitted augmented Cox procedure for studies
in which follow-up time and regression covariates are accurately
observed but the gold-standard failure indicator is available only for
an internal validation sample. The point-estimating equation has
established augmented inverse-probability ancestry. The principal
methodological contribution is instead a coherent first-order treatment
of the cross-fitted pseudo-event Cox score: empirical risk-set
linearization produces the same score contribution obtained
independently by projecting the full-data Cox influence function through
the validation mechanism. That agreement organizes analytic inference,
while the exact product-form nuisance drift supplies double-robust
identification and Neyman orthogonality and the influence representation
leads directly to validation design.

OVAC brings classical augmented survival estimation into a computational
regime in which the event bridge and, when needed, the validation model
can be learned flexibly out of fold. Cross-fitting and the product-rate
remainder protect root-n target inference from first-order nuisance
error at the intersection model. The coarsening projection clarifies the
efficiency target, while the validation allocation result translates the
same influence structure into a design criterion that favors records
with both classification uncertainty and leverage for the Cox contrast
of interest. The \(G_{0}\) diagnostics sharpen the interpretation of the
risk-set result. In the present DGP, \(G_{0}\) alone has roughly 78\%-81\% of
the full-SE magnitude, yet deleting it changes the net analytic SE by
less than 0.3\% because its variance is almost canceled by twice its
covariance with the event-score contribution. The small ablation is
therefore not evidence that \(G_{0}\) is a negligible first-order term. It is
evidence that one benign calibration regime happens to produce a strong
covariance cancellation, whose size and sign are not protected by the
theory.

The scope of the result is intentionally specific. The full-data Cox
model must itself be scientifically appropriate, and endpoint correction
does not address nonproportional hazards, uncontrolled confounding,
dependent censoring, or an unsuitable estimand. The core theory also
assumes accurately observed follow-up times and regression covariates.
When those quantities are error-prone, the at-risk process is no longer
fully observed and the pseudo-event simplification is insufficient;
generalized raking, multiple imputation, likelihood, SIMEX, and broader
semiparametric measurement-error methods remain relevant.

The most direct application is a large cohort with limited adjudication
resources. Investigators can validate an informative subset, learn an
event bridge from phase-I signals, and analyze the full cohort with a
score that remains consistent if either the bridge or validation
mechanism is correct. When both are estimated well, augmentation
recovers information from nonvalidated records and can substantially
improve efficiency over IPW. The allocation theorem then provides a
principled way to target later validation waves toward records that are
difficult to classify and influential for the coefficient of interest.

Several extensions require additional theory. Event-time error would
require joint recovery of the at-risk and counting processes.
Competing-risk extensions should treat the cause bridge as multinomial
and derive joint influence functions across causes. Recurrent-event and
multi-state settings replace the single hidden indicator by a sequence
of potentially misclassified transition increments. Highly adaptive
nuisance learning also motivates comparison of the score-root estimator
with one-step or targeted implementations. Finally, rare-event settings
deserve dedicated study because the optimal validation rule can become
highly concentrated when the true event probability is small for most
records.

The central structural lesson is that endpoint validation problems
should be matched to the part of the counting process that is actually
uncertain. When follow-up time and covariates are reliable and only the
failure label is hidden, the Cox score admits a particularly simple
augmented representation. Exploiting that structure yields a transparent
route from identification to cross-fitted inference and validation
design without invoking a broader error model than the data require.

\section{Acknowledgments and
declarations}

\textbf{Funding.} This methodological study received no external
funding.

\textbf{Conflicts of interest.} None declared.

\textbf{Data availability.} No human-subject data were analyzed. All
numerical results are based on synthetic Monte Carlo data generated from
the mechanism specified in Section 9.

Code availability. Versioned R research code implements cross-fitted
OVAC estimation, direct pseudo-event risk-set linearization, the
bridge-stabilized diagnostic, Monte Carlo calibration, and
validation-design calculations. The primary 1,000-replication variance
calibration and \(G_{0}\) ablation are reproducible in R, and an independently
coded Python implementation provides a cross-language verification of
the analytic variance quantities. The component diagnostic separates the
transformed event-score and \(G_{0}\) contributions and reports their
individual SE magnitudes and correlation. Seed-controlled scripts and
numerical outputs are intended to accompany the preprint and journal
submission.

\appendix
\section{Full mathematical proofs}

This appendix gives complete proofs of every formal result in the main
text.  The arguments use standard counting-process facts for the Cox
model and standard empirical-process facts for the at-risk classes, but
all algebra specific to OVAC, cross-fitting, risk-set linearization,
coarsening, and validation design is shown explicitly.

Throughout Appendix A, write
\[
  \bar{X}_0(t)=\bar{X}(t,\beta_0),\qquad
  s_0(t)=s^{(0)}(t,\beta_0),\qquad
  c_0(t,X)=X-\bar{X}_0(t),
\]
and
\[
  w_0(X)=\exp(\beta_0^\top X),\qquad
  f_t(O)=Y(t)w_0(X)c_0(t,X).
\]
For a measurable function \(g\), let \(Pg=E\{g(O)\}\),
\(P_ng=n^{-1}\sum_{i=1}^n g(O_i)\), and
\(\mathbb G_ng=\sqrt n(P_n-P)g\).  For fold \(k\), \(P_{n,k}\) denotes
the empirical measure over the held-out observations in that fold.
All vector norms below are Euclidean and all matrix norms are operator
norms.  The constants denoted by \(C\) may change from line to line.

\subsection{Preliminary identities}

\subsubsection{Mean and boundedness of the intersection-model pseudo-event}

At the intersection model define
\[
 D_0
 =p_0(V)+\frac{R}{\pi_0(V)}\{\Delta-p_0(V)\}.
\]
By (6)--(8),
\[
 E(D_0\mid V)
 =p_0(V)+\frac{E(R\mid V)}{\pi_0(V)}
   \{E(\Delta\mid V,R=1)-p_0(V)\}
 =p_0(V).
\tag{A.1}
\]
Hence, for every integrable phase-I-measurable \(a(V)\),
\[
 E\{D_0a(V)\}=E\{p_0(V)a(V)\}
             =E\{\Delta a(V)\}.
\tag{A.2}
\]
Because \(0\le p_0\le1\), \(0\le R,\Delta\le1\), and
\(\pi_0\ge\varepsilon\),
\[
 |D_0|
 \le 1+\varepsilon^{-1}.
\tag{A.3}
\]
The same type of bound holds with probability tending to one for every
cross-fitted \(\widetilde D_i\) under (A7).

\subsubsection{Exact risk-set ratio identity}

At \(\beta_0\),
\[
 \bar{X}_n(t,\beta_0)-\bar{X}_0(t)
 =
 \frac{
  P_n\{Y(t)w_0(X)[X-\bar{X}_0(t)]\}
 }{
  S_n^{(0)}(t,\beta_0)
 }
 =
 \frac{P_nf_t}{S_n^{(0)}(t,\beta_0)}.
\tag{A.4}
\]
This is an exact algebraic identity, not an approximation: subtract
\(\bar{X}_0(t)S_n^{(0)}(t,\beta_0)\) from
\(S_n^{(1)}(t,\beta_0)\) and divide by
\(S_n^{(0)}(t,\beta_0)\).

Under (A1)--(A4), the classes
\[
 \{Y(t)w_0(X):0\le t\le\tau\},\qquad
 \{f_t:0\le t\le\tau\}
\]
are Donsker under bounded \(X\); under the stated exponential-moment
alternative the same conclusions follow by the usual localization
argument used for the Cox score.  Therefore
\[
 \sup_{t\le\tau}
 |S_n^{(0)}(t,\beta_0)-s_0(t)|=O_p(n^{-1/2}),
\qquad
 \sup_{t\le\tau}\|P_nf_t\|=O_p(n^{-1/2}).
\tag{A.5}
\]
Since \(\inf_{t\le\tau}s_0(t)>0\), (A.5) also implies
\[
 \sup_{t\le\tau}
 \left|
 \frac{1}{S_n^{(0)}(t,\beta_0)}-\frac{1}{s_0(t)}
 \right|
 =O_p(n^{-1/2})
\tag{A.6}
\]
and, by (A.4),
\[
 \sup_{t\le\tau}
 \|\bar{X}_n(t,\beta_0)-\bar{X}_0(t)\|
 =O_p(n^{-1/2}).
\tag{A.7}
\]

\subsection{Proof of Lemma 1}

\begin{proof}
Let \(m_0(V)=E(H\mid V)\).  Conditional on \(V\), the working functions
\(m(V)\) and \(\pi(V)\) are constants.  By \(R\perp H\mid V\),
\[
\begin{aligned}
 E\{\mathcal A(H;m,\pi)\mid V\}
 &=
 m(V)+\frac{1}{\pi(V)}
 E\{R[H-m(V)]\mid V\}\\
 &=
 m(V)+\frac{1}{\pi(V)}
 E(R\mid V)
 E\{H-m(V)\mid V\}\\
 &=
 m(V)+
 \frac{\pi_0(V)}{\pi(V)}
 \{m_0(V)-m(V)\}.
\end{aligned}
\]
The conditional expectation exists by the assumed integrability and
positivity of \(\pi\).  This is the claimed identity.
\end{proof}

\subsection{Proof of Proposition 1}

\begin{proof}
Apply Lemma 1 with \(H=\Delta\), \(m=p\), and
\(m_0=p_0\).  Conditional on \(V\),
\[
\begin{aligned}
 E\{D(p,\pi)\mid V\}
 &=
 p(V)+\frac{\pi_0(V)}{\pi(V)}
       \{p_0(V)-p(V)\}\\
 &=
 p_0(V)+
 \left\{1-\frac{\pi_0(V)}{\pi(V)}\right\}
 \{p(V)-p_0(V)\}.
\end{aligned}
\tag{A.8}
\]
Because \(c_\beta(U,X)\) is \(V\)-measurable,
\[
\begin{aligned}
 \Psi(\beta;p,\pi)
 &=
 E\!\left[
 c_\beta(U,X)E\{D(p,\pi)\mid V\}
 \right]\\
 &=
 E\{c_\beta(U,X)p_0(V)\}\\
 &\quad+
 E\!\left[
 c_\beta(U,X)
 \left\{1-\frac{\pi_0(V)}{\pi(V)}\right\}
 \{p(V)-p_0(V)\}
 \right].
\end{aligned}
\]
By iterated expectation,
\[
 E\{c_\beta(U,X)p_0(V)\}
 =
 E[E\{\Delta\mid V\}c_\beta(U,X)]
 =
 E\{\Delta c_\beta(U,X)\}
 =
 \Psi_F(\beta).
\]
Subtracting \(\Psi_F(\beta)\) proves (16).
\end{proof}

\subsection{Proof of Proposition 2}

\begin{proof}
If \(p=p_0\) almost surely, the last factor in (16) is zero almost
surely.  If \(\pi=\pi_0\) almost surely, the middle factor in (16) is
zero almost surely.  Therefore in either model
\[
 \Psi(\beta;p,\pi)=\Psi_F(\beta)
 \qquad\text{for every }\beta.
\]
Since the full-data score identifies \(\beta_0\), the same
\(\beta_0\) is the unique local root of the OVAC population score.
\end{proof}

\subsection{Proof of Lemma 2}

\begin{proof}
It is useful to prove the stronger joint-direction statement.  Let
\[
 p_r=p_0+r h_p,\qquad
 \pi_r=\pi_0+r h_\pi
\]
for \(r\) in a neighborhood of zero small enough that
\(\pi_r\) remains positive.  Proposition 1 gives
\[
\begin{aligned}
 \Psi(\beta_0;p_r,\pi_r)-\Psi_F(\beta_0)
 &=
 E\!\left[
 c_0(U,X)
 \frac{\pi_r(V)-\pi_0(V)}{\pi_r(V)}
 \{p_r(V)-p_0(V)\}
 \right]\\
 &=
 r^2 E\!\left[
 c_0(U,X)
 \frac{h_\pi(V)h_p(V)}
      {\pi_0(V)+r h_\pi(V)}
 \right].
\end{aligned}
\tag{A.9}
\]
Under the square-integrability assumptions and positivity, the
integrand divided by \(r^2\) is dominated in a neighborhood of zero.
Consequently the right-hand side is \(O(r^2)\), so its derivative at
zero is zero.  In particular the derivatives in the separate \(p\) and
\(\pi\) directions are zero, proving Neyman orthogonality.
\end{proof}

\subsection{Proof of the product-rate bound}

\begin{proof}
From Proposition 1 and
\(1-\pi_0/\widehat\pi=(\widehat\pi-\pi_0)/\widehat\pi\),
\[
 \Psi(\beta_0;\widehat p,\widehat\pi)
 =
 E\!\left[
 c_0(U,X)
 \frac{\widehat\pi(V)-\pi_0(V)}{\widehat\pi(V)}
 \{\widehat p(V)-p_0(V)\}
 \right],
\]
because \(\Psi_F(\beta_0)=0\).  On the event
\(\inf_V\widehat\pi(V)\ge\varepsilon/2\), and writing
\(C_c=\sup\|c_0(U,X)\|\),
\[
\begin{aligned}
 \|\Psi(\beta_0;\widehat p,\widehat\pi)\|
 &\le
 \frac{2C_c}{\varepsilon}
 E\!\left[
 |\widehat\pi-\pi_0|\,|\widehat p-p_0|
 \right]\\
 &\le
 \frac{2C_c}{\varepsilon}
 \|\widehat\pi-\pi_0\|_2
 \|\widehat p-p_0\|_2,
\end{aligned}
\]
where the last step is Cauchy--Schwarz.  This proves the bound in \eqref{eq:17}, with
\(C=2C_c/\varepsilon\).  Under (A7) the right-hand side is
\(o_p(n^{-1/2})\).
\end{proof}

\subsection{Proof of Lemma 3}

\begin{proof}
Write \(s^{(k)}_\beta(t)=s^{(k)}(t,\beta)\) and
\(\bar{X}_\beta(t)=s^{(1)}_\beta(t)/s^{(0)}_\beta(t)\).  Differentiation
under the expectation is justified by (A3)--(A4).  Componentwise,
\[
 \frac{\partial s^{(0)}_\beta(t)}{\partial\beta}
 =s^{(1)}_\beta(t),
 \qquad
 \frac{\partial s^{(1)}_\beta(t)}{\partial\beta^\top}
 =s^{(2)}_\beta(t).
\]
The quotient rule therefore gives
\[
\begin{aligned}
 \frac{\partial\bar{X}_\beta(t)}{\partial\beta^\top}
 &=
 \frac{s^{(2)}_\beta(t)}{s^{(0)}_\beta(t)}
 -
 \frac{s^{(1)}_\beta(t)s^{(1)}_\beta(t)^\top}
      {s^{(0)}_\beta(t)^2}\\
 &=v(t,\beta).
\end{aligned}
\tag{A.10}
\]
Since
\(\Psi_F(\beta)=E[\Delta\{X-\bar{X}_\beta(U)\}]\),
\[
 -\frac{\partial\Psi_F(\beta)}
        {\partial\beta^\top}\Big|_{\beta=\beta_0}
 =
 E\{\Delta v(U,\beta_0)\}.
\tag{A.11}
\]
Under the Cox multiplicative-intensity model,
\[
 E\{dN(t)\mid\mathcal F_{t-}\}
 =
 Y(t)w_0(X)d\Lambda_0(t).
\]
Hence Tonelli's theorem and iterated expectation imply
\[
\begin{aligned}
 E\{\Delta v(U,\beta_0)\}
 &=
 E\int_0^\tau v(t,\beta_0)dN(t)\\
 &=
 \int_0^\tau
 E\{Y(t)w_0(X)v(t,\beta_0)\}
 d\Lambda_0(t)\\
 &=
 \int_0^\tau
 s_0(t)v(t,\beta_0)d\Lambda_0(t),
\end{aligned}
\]
because \(v(t,\beta_0)\) is deterministic for fixed \(t\).  This proves
(18).
\end{proof}

\subsection{Proof of Theorem 1}

\begin{proof}
Let \(B\) be the compact neighborhood in the theorem and define
\[
 c_{n,\beta}(U,X)
 =
 X-\bar{X}_n(U,\beta),\qquad
 c_\beta(U,X)
 =
 X-\bar{X}(U,\beta).
\]
For notational clarity, write
\(\widetilde D_i=D(\widehat p_{-k(i)},\widehat\pi_{-k(i)};O_i)\).
Under the stated boundedness and positivity,
\[
 \max_i|\widetilde D_i|\le C
\]
with probability tending to one.

We show
\[
 \sup_{\beta\in B}
 \|U_n(\beta)-\Psi_F(\beta)\|=o_p(1).
\tag{A.12}
\]
Decompose
\[
\begin{aligned}
 U_n(\beta)-\Psi_F(\beta)
 &=
 P_n\{\widetilde D[c_{n,\beta}-c_\beta]\}\\
 &\quad+
 \{P_n(\widetilde D c_\beta)-P(\widetilde D c_\beta)\}\\
 &\quad+
 \{\Psi(\beta;\widehat p,\widehat\pi)-\Psi_F(\beta)\},
\end{aligned}
\tag{A.13}
\]
where the second and third terms are interpreted foldwise because the
working nuisance functions differ across folds.

For the first term, uniform risk-set convergence on \(B\) gives
\[
 \sup_{\beta\in B}
 \sup_{t\le\tau}
 \|\bar{X}_n(t,\beta)-\bar{X}(t,\beta)\|=o_p(1).
\]
Since \(P_n|\widetilde D|=O_p(1)\), the first line of (A.13) is
\(o_p(1)\) uniformly on \(B\).

For the second line, condition on the training sample of a fixed fold.
The held-out observations are i.i.d. and the fitted nuisance functions
are fixed.  The summands are uniformly bounded by an integrable
envelope, so a conditional uniform law of large numbers applies to
\(\{\widetilde D c_\beta:\beta\in B\}\).  Because the number of folds is
fixed and each fold has asymptotically positive sample fraction, summing
the foldwise \(o_p(1)\) terms remains \(o_p(1)\).

For the population bias in the third line, Proposition 1 yields
\[
 \|\Psi(\beta;\widehat p,\widehat\pi)-\Psi_F(\beta)\|
 \le
 C
 \|\widehat p-p_0\|_2
 \|\widehat\pi-\pi_0\|_2
\tag{A.14}
\]
whenever both errors are used.  Under the union-model assumptions we
need less.  If \(\|\widehat p-p_0\|_2=o_p(1)\), then
\(\|\widehat\pi-\pi_0\|_2=O_p(1)\) because both probabilities are
bounded and \(\widehat\pi\ge\varepsilon/2\); therefore (A.14) is
\(o_p(1)\).  The same argument applies with the roles reversed if
\(\widehat\pi\to\pi_0\), because \(0\le\widehat p,p_0\le1\).  The bound
is uniform in \(\beta\in B\) because \(c_\beta\) has a common
integrable envelope there.  Thus (A.12) holds.

Since \(\Psi_F\) is continuous, has the unique zero \(\beta_0\) on
\(B\), and is separated from zero outside every neighborhood of
\(\beta_0\), the usual Z-estimator argument is elementary.  For any
\(\delta>0\), compactness gives
\[
 \eta_\delta
 =
 \inf_{\beta\in B:\|\beta-\beta_0\|\ge\delta}
 \|\Psi_F(\beta)\|>0.
\]
On the event
\(\sup_{\beta\in B}\|U_n(\beta)-\Psi_F(\beta)\|<\eta_\delta/3\) and
\(\|U_n(\widehat\beta)\|<\eta_\delta/3\), if
\(\|\widehat\beta-\beta_0\|\ge\delta\) then
\[
 \eta_\delta
 \le \|\Psi_F(\widehat\beta)\|
 \le
 \|\Psi_F(\widehat\beta)-U_n(\widehat\beta)\|
 +\|U_n(\widehat\beta)\|
 <\frac{2}{3}\eta_\delta,
\]
a contradiction.  Both events have probability tending to one.
Therefore
\(\widehat\beta\overset p\longrightarrow\beta_0\).
\end{proof}

\subsection{Auxiliary lemma for Theorem 2: cross-fitted nuisance replacement}

\textbf{Lemma A.1.}
Under (A1)--(A8),
\[
 U_n(\beta_0,\widehat\eta)
 -
 U_n(\beta_0,\eta_0)
 =
 o_p(n^{-1/2}),
\tag{A.15}
\]
where \(\eta_0=(p_0,\pi_0)\) and the score evaluated at \(\eta_0\) uses \(D_0\) but the
same empirical risk-set mean \(\bar{X}_n\).

\begin{proof}
Set
\[
 \delta_i
 =
 D(\widehat p_{-k(i)},\widehat\pi_{-k(i)};O_i)-D_0(O_i),
 \qquad
 r_n(t)=\bar{X}_n(t,\beta_0)-\bar{X}_0(t).
\]
Then
\[
 U_n(\beta_0,\widehat\eta)-U_n(\beta_0,\eta_0)
 =
 P_n\{\delta c_0(U,X)\}
 -
 P_n\{\delta r_n(U)\}.
\tag{A.16}
\]

First consider \(P_n\{\delta c_0\}\).  For fold \(k\), let
\(\delta_k(O)\) denote the difference generated by nuisance fits trained
outside that fold.  Conditional on the training sample,
\[
 P_{n,k}(\delta_kc_0)
 =
 (P_{n,k}-P)(\delta_kc_0)+P(\delta_kc_0).
\tag{A.17}
\]
The pseudo-event map is locally Lipschitz in \((p,\pi)\) under
\(\pi,\pi_0\ge\varepsilon/2\).  Indeed, for arbitrary \(p,\pi\),
\[
\begin{aligned}
 D(p,\pi)-D(p_0,\pi_0)
 &=
 (p-p_0)\left(1-\frac{R}{\pi_0}\right)\\
 &\quad+
 R(\Delta-p)
 \left(\frac1\pi-\frac1{\pi_0}\right).
\end{aligned}
\]
Because \(R,\Delta,p,p_0\) are bounded and
\[
 \left|\frac1\pi-\frac1{\pi_0}\right|
 =
 \frac{|\pi-\pi_0|}{\pi\pi_0}
 \le \frac{4}{\varepsilon^2}|\pi-\pi_0|,
\]
we obtain
\[
 |\delta_k|
 \le
 C\left(
 |\widehat p_k-p_0|
 +
 |\widehat\pi_k-\pi_0|
 \right).
\tag{A.18}
\]  Thus
\[
 \|\delta_k\|_2=o_p(1).
\tag{A.19}
\]
Conditional on the training sample, the first term in (A.17) has mean
zero and variance at most
\[
 \frac{1}{n_k}
 P\|\delta_kc_0\|^2
 \le
 \frac{C}{n_k}\|\delta_k\|_2^2
 =
 o_p(n_k^{-1}).
\]
Chebyshev's inequality therefore gives
\[
 (P_{n,k}-P)(\delta_kc_0)=o_p(n^{-1/2}).
\tag{A.20}
\]
For the second term, Proposition 1 at \(\beta_0\) gives exactly
\[
 P(\delta_kc_0)
 =
 E\!\left[
 c_0
 \frac{\widehat\pi_k-\pi_0}{\widehat\pi_k}
 (\widehat p_k-p_0)
 \right],
\]
so by \eqref{eq:17}
\[
 P(\delta_kc_0)
 =
 O_p\!\left(
 \|\widehat p_k-p_0\|_2
 \|\widehat\pi_k-\pi_0\|_2
 \right)
 =
 o_p(n^{-1/2}).
\tag{A.21}
\]
Because \(K\) is fixed, the fold-weighted sum of (A.20)--(A.21) is
\(o_p(n^{-1/2})\).

For the second term in (A.16), (A.7) and Cauchy--Schwarz give
\[
\begin{aligned}
 \|P_n\{\delta r_n(U)\}\|
 &\le
 \sup_{t\le\tau}\|r_n(t)\|\,P_n|\delta|\\
 &=
 O_p(n^{-1/2})\,o_p(1)
 =
 o_p(n^{-1/2}),
\end{aligned}
\tag{A.22}
\]
where \(P_n|\delta|=o_p(1)\) follows foldwise from (A.18)--(A.19).
Combining the two parts proves (A.15).
\end{proof}

\subsection{Auxiliary lemma for Theorem 2: risk-set linearization at the true nuisance functions}

\textbf{Lemma A.2.}
At the intersection model,
\[
 \sqrt n\,U_n(\beta_0,\eta_0)
 =
 \frac{1}{\sqrt n}\sum_{i=1}^n
 \left[
 D_{0i}c_0(U_i,X_i)-G_0(U_i,X_i)
 \right]+o_p(1).
\tag{A.23}
\]

\begin{proof}
The score evaluated at \(\eta_0\) is
\[
 U_n(\beta_0,\eta_0)
 =
 P_n\{D_0c_0(U,X)\}
 -
 A_n,
\tag{A.24}
\]
where, by (A.4),
\[
 A_n
 =
 P_n\!\left[
 D_0
 \frac{P_nf_U}{S_n^{(0)}(U,\beta_0)}
 \right].
\tag{A.25}
\]
Define
\[
 B_n
 =
 P_n\!\left[
 D_0
 \frac{P_nf_U}{s_0(U)}
 \right].
\tag{A.26}
\]
By (A.5)--(A.6), (A.3), and \(P_n|D_0|=O_p(1)\),
\[
\begin{aligned}
 |A_n-B_n|
 &\le
 P_n|D_0|
 \sup_t\|P_nf_t\|
 \sup_t
 \left|
 \frac{1}{S_n^{(0)}(t,\beta_0)}-\frac{1}{s_0(t)}
 \right|\\
 &=O_p(n^{-1}),
\end{aligned}
\tag{A.27}
\]
so it suffices to linearize \(B_n\).

Write
\[
 B_n=\frac1{n^2}\sum_{i=1}^n\sum_{j=1}^n h(O_i,O_j),
\tag{A.28}
\]
with
\[
 h(o_i,o_j)
 =
 \frac{D_{0i}}{s_0(U_i)}
 Y_j(U_i)w_0(X_j)c_0(U_i,X_j).
\tag{A.29}
\]
The two first-order projections of this non-symmetric kernel have a
simple form.  For fixed \(o_i\),
\[
\begin{aligned}
 E_j\{h(o_i,O_j)\}
 &=
 \frac{D_{0i}}{s_0(U_i)}
 E\{Y(U_i)w_0(X)c_0(U_i,X)\}\\
 &=
 \frac{D_{0i}}{s_0(U_i)}
 \{s^{(1)}(U_i,\beta_0)-s_0(U_i)\bar{X}_0(U_i)\}\\
 &=0.
\end{aligned}
\tag{A.30}
\]
For fixed \(o_j\), use (A.2) and the Cox compensator identity:
\[
 E\{D_0 I(U\in dt)\}
 =
 E\{\Delta I(U\in dt)\}
 =
 E\{dN(t)\}
 =
 s_0(t)d\Lambda_0(t).
\tag{A.31}
\]
Therefore
\[
\begin{aligned}
 E_i\{h(O_i,o_j)\}
 &=
 w_0(X_j)
 \int_0^{U_j}
 c_0(t,X_j)
 \frac{E\{D_0 I(U\in dt)\}}{s_0(t)}\\
 &=
 w_0(X_j)
 \int_0^{U_j}c_0(t,X_j)d\Lambda_0(t)\\
 &=G_0(U_j,X_j).
\end{aligned}
\tag{A.32}
\]
Also \(EG_0=0\), since
\[
 EG_0
 =
 \int_0^\tau
 E\{Y(t)w_0(X)c_0(t,X)\}d\Lambda_0(t)
 =0.
\tag{A.33}
\]

To make the V-statistic reduction explicit, separate the diagonal:
\[
 B_n
 =
 \frac{n-1}{n}U_n^{(2)}+\frac1{n^2}\sum_{i=1}^n h(O_i,O_i),
\]
where \(U_n^{(2)}\) is the order-two U-statistic based on the symmetrized
kernel
\[
 h_s(o_1,o_2)=\frac12\{h(o_1,o_2)+h(o_2,o_1)\}.
\]
The diagonal term is \(O_p(n^{-1})\) under the square-integrability
implied by (A1)--(A4).  The first Hoeffding projection of \(h_s\) is,
by (A.30)--(A.32),
\[
 h_1(o)
 =
 E\{h_s(o,O')\}
 =
 \frac12G_0(o).
\]
The Hoeffding decomposition gives
\[
 U_n^{(2)}
 =
 2P_nh_1+R_n^{(2)}
 =
 P_nG_0+R_n^{(2)},
\tag{A.34}
\]
where the degenerate remainder satisfies
\(E\|R_n^{(2)}\|^2=O(n^{-2})\), hence
\(R_n^{(2)}=O_p(n^{-1})\).  Consequently
\[
 B_n=P_nG_0+O_p(n^{-1}).
\tag{A.35}
\]
Together with (A.27),
\[
 A_n=P_nG_0+o_p(n^{-1/2}).
\tag{A.36}
\]
Substituting (A.36) into (A.24) proves (A.23).
\end{proof}

\subsection{Proof of Theorem 2}

\begin{proof}
By Theorem 1, under (A1)--(A8) both nuisances are consistent and hence
\(\widehat\beta\to_p\beta_0\).  Define the negative score derivative
\[
 J_n(\beta)
 =
 -\frac{\partial U_n(\beta)}{\partial\beta^\top}
 =
 P_n\{\widetilde D\,v_n(U,\beta)\},
\tag{A.37}
\]
where
\[
 v_n(t,\beta)
 =
 \frac{S_n^{(2)}(t,\beta)}{S_n^{(0)}(t,\beta)}
 -
 \bar{X}_n(t,\beta)^{\otimes2}.
\]
Uniform convergence of the risk-set moments implies
\[
 \sup_{\beta\in B,t\le\tau}
 \|v_n(t,\beta)-v(t,\beta)\|=o_p(1).
\tag{A.38}
\]
At \(\beta_0\), write
\[
\begin{aligned}
 J_n(\beta_0)-I_0
 &=
 P_n\{(\widetilde D-D_0)v(U,\beta_0)\}\\
 &\quad+
 P_n\{D_0[v_n(U,\beta_0)-v(U,\beta_0)]\}\\
 &\quad+
 \{P_n[D_0v(U,\beta_0)]-E[\Delta v(U,\beta_0)]\}.
\end{aligned}
\tag{A.39}
\]
The first term is \(o_p(1)\) by (A.18)--(A.19), the second is \(o_p(1)\)
by (A.38) and bounded \(D_0\), and the third is \(o_p(1)\) by the law of
large numbers and (A.2).  Stochastic equicontinuity in \(\beta\), plus
\(\widehat\beta\to_p\beta_0\), therefore gives
\[
 J_n(\widetilde\beta)\to_p I_0
\tag{A.40}
\]
for every \(\widetilde\beta\) on the line segment joining
\(\widehat\beta\) and \(\beta_0\).

The approximate-root condition and the multivariate mean-value
expansion yield
\[
 U_n(\widehat\beta)
 =
 U_n(\beta_0)
 -
 J_n(\widetilde\beta)(\widehat\beta-\beta_0),
\]
hence
\[
 \sqrt n(\widehat\beta-\beta_0)
 =
 J_n(\widetilde\beta)^{-1}
 \sqrt n\,U_n(\beta_0)
 +o_p(1).
\tag{A.41}
\]
By Lemma A.1,
\[
 \sqrt n\{
 U_n(\beta_0,\widehat\eta)-U_n(\beta_0,\eta_0)
 \}=o_p(1),
\tag{A.42}
\]
and by Lemma A.2,
\[
 \sqrt n\,U_n(\beta_0,\eta_0)
 =
 \frac1{\sqrt n}\sum_{i=1}^n
 \xi_0(O_i)+o_p(1),
\quad
 \xi_0(O)=D_0c_0(U,X)-G_0(U,X).
\tag{A.43}
\]
Combining (A.40)--(A.43) and Slutsky's theorem gives
\[
 \sqrt n(\widehat\beta-\beta_0)
 =
 I_0^{-1}
 \frac1{\sqrt n}\sum_{i=1}^n\xi_0(O_i)+o_p(1),
\]
which is (19).
\end{proof}

\subsection{Proof of Corollary 1}

\begin{proof}
Equation (A.2) and the population Cox score imply
\[
 E\{D_0c_0(U,X)\}
 =
 E\{\Delta c_0(U,X)\}=0.
\]
Equation (A.33) gives \(EG_0=0\), so \(E\xi_0=0\).  Under
(A1)--(A6), \(\xi_0\) has finite second moment.  Therefore the
multivariate central limit theorem gives
\[
 \frac1{\sqrt n}\sum_{i=1}^n\xi_0(O_i)
 \Rightarrow
 N_p\{0,E(\xi_0\xi_0^\top)\}.
\]
Theorem 2 and Slutsky's theorem imply
\[
 \sqrt n(\widehat\beta-\beta_0)
 \Rightarrow
 N_p\!\left(
 0,\,
 I_0^{-1}E(\xi_0\xi_0^\top)I_0^{-1}
 \right).
\]
Theorem 3 below identifies
\(I_0^{-1}\xi_0=\phi_{\mathrm{eff}}\), so the covariance is
\(V_{\mathrm{eff}}=E(\phi_{\mathrm{eff}}\phi_{\mathrm{eff}}^\top)\),
proving (20).
\end{proof}

\subsection{Proof of Proposition 3}

\begin{proof}
We derive the efficient score rather than merely quote the Cox result.
Under the multiplicative-intensity formulation, the failure-process
martingale is
\[
 M(t)
 =
 N(t)-\int_0^tY(u)w_0(X)d\Lambda_0(u).
\]
A regular parametric submodel
\(\beta_\epsilon=\beta_0+\epsilon b\) has score in direction \(b\)
\[
 S_\beta(b)
 =
 b^\top\int_0^\tau X\,dM(t).
\tag{A.44}
\]
A regular baseline-hazard submodel
\(d\Lambda_\epsilon(t)=\{1+\epsilon a(t)+o(\epsilon)\}d\Lambda_0(t)\)
has nuisance score
\[
 S_\Lambda(a)=\int_0^\tau a(t)\,dM(t).
\tag{A.45}
\]
The nuisance tangent spaces for the marginal distribution of \(X\) and
for the independent censoring law are orthogonal to the failure
martingale score: the first because
\(E\{\int X\,dM\mid X\}=0\), and the second by orthogonality of the
failure and censoring martingales under conditional independent
censoring.  Hence only the baseline-hazard tangent needs to be projected
out of (A.44).

For any square-integrable \(a\), the martingale isometry gives
\[
\begin{aligned}
 E\!\left[
 \left\{\int_0^\tau X\,dM(t)\right\}
 \left\{\int_0^\tau a(t)\,dM(t)\right\}
 \right]
 &=
 \int_0^\tau s^{(1)}(t,\beta_0)a(t)d\Lambda_0(t).
\end{aligned}
\tag{A.46}
\]
Similarly,
\[
\begin{aligned}
 E\!\left[
 \left\{\int_0^\tau \bar{X}_0(t)\,dM(t)\right\}
 \left\{\int_0^\tau a(t)\,dM(t)\right\}
 \right]
 &=
 \int_0^\tau
 \bar{X}_0(t)s_0(t)a(t)d\Lambda_0(t)\\
 &=
 \int_0^\tau
 s^{(1)}(t,\beta_0)a(t)d\Lambda_0(t).
\end{aligned}
\tag{A.47}
\]
Therefore
\[
 S_{\mathrm{eff}}
 =
 \int_0^\tau c_0(t,X)dM(t)
\tag{A.48}
\]
is orthogonal to every baseline-hazard nuisance score, and
\(\int\bar{X}_0\,dM\) is exactly the projection removed from the raw
\(\beta\)-score.

Its covariance is
\[
\begin{aligned}
 E(S_{\mathrm{eff}}S_{\mathrm{eff}}^\top)
 &=
 E\int_0^\tau
 c_0(t,X)^{\otimes2}
 Y(t)w_0(X)d\Lambda_0(t)\\
 &=
 \int_0^\tau
 \left[
 s^{(2)}(t,\beta_0)
 -
 \frac{s^{(1)}(t,\beta_0)s^{(1)}(t,\beta_0)^\top}
      {s_0(t)}
 \right]d\Lambda_0(t)\\
 &=
 \int_0^\tau s_0(t)v(t,\beta_0)d\Lambda_0(t)
 =
 I_0,
\end{aligned}
\tag{A.49}
\]
using Lemma 3.  Since \(I_0\) is nonsingular, the full-data canonical
gradient is
\[
 \phi_F=I_0^{-1}S_{\mathrm{eff}},
\]
which proves (22).

Finally, \(N\) has the single jump \(\Delta\) at \(U\), so
\[
 \int_0^\tau c_0(t,X)dN(t)
 =
 \Delta c_0(U,X).
\]
The compensator part is
\[
 \int_0^\tau
 c_0(t,X)Y(t)w_0(X)d\Lambda_0(t)
 =
 w_0(X)\int_0^Uc_0(t,X)d\Lambda_0(t)
 =
 G_0(U,X).
\]
Substitution into (22) yields (23)--(24).
\end{proof}

\subsection{Proof of Lemma 4}

\begin{proof}
Let \(F=(V,H)\), \(O=(V,R,RH)\), and
\(h_\phi(V)=E\{\phi_F(F)\mid V\}\).  Define
\[
 \phi_O(O)
 =
 h_\phi(V)
 +
 \frac{R}{\pi_0(V)}
 \{\phi_F(F)-h_\phi(V)\}.
\tag{A.50}
\]
First, \(E\phi_O=0\), because conditional on \(V\),
\[
 E(\phi_O\mid V)
 =
 h_\phi(V)
 +
 \frac{E(R\mid V)}{\pi_0(V)}
 E\{\phi_F-h_\phi(V)\mid V,R=1\}
 =
 h_\phi(V),
\]
where \(R\perp H\mid V\), and \(Eh_\phi=E\phi_F=0\).

We next verify the pathwise-derivative equation.  Let \(S_F(F)\) be an
arbitrary regular score perturbing the full-data law while the
validation law is held fixed.  The corresponding observed-data score is
\(S_O(O)=E\{S_F(F)\mid O\}\).  Since \(\phi_O\) is \(O\)-measurable,
\[
 E(\phi_OS_O)
 =
 E\{\phi_O S_F(F)\}.
\tag{A.51}
\]
Conditional on the full data \(F\), only \(R\) is random, and
\[
 E(\phi_O\mid F)
 =
 h_\phi(V)
 +
 \frac{E(R\mid F)}{\pi_0(V)}
 \{\phi_F-h_\phi(V)\}
 =
 \phi_F,
\tag{A.52}
\]
again using \(E(R\mid F)=E(R\mid V)=\pi_0(V)\).  Consequently
\[
 E(\phi_OS_O)
 =
 E(\phi_FS_F),
\tag{A.53}
\]
which is the full-data pathwise derivative represented by the canonical
gradient \(\phi_F\).

It remains to show orthogonality to perturbations of the validation
mechanism.  Every regular validation score can be represented, up to
closure, by
\[
 S_R(O)=(R-\pi_0(V))a(V)
\]
with a square-integrable scaling absorbed into \(a\).  Conditional on
\(V\),
\[
\begin{aligned}
 E\{\phi_O(R-\pi_0)\mid V\}
 &=
 h_\phi(V)E(R-\pi_0\mid V)\\
 &\quad+
 E\!\left[
 \{\phi_F-h_\phi(V)\}\mid V
 \right]
 \frac{E\{R(R-\pi_0)\mid V\}}{\pi_0(V)}\\
 &=0.
\end{aligned}
\tag{A.54}
\]
Thus \(E(\phi_OS_R)=0\) for every validation nuisance direction.
Equations (A.53)--(A.54) show that \(\phi_O\) represents the target
derivative on the observed-data tangent space and is orthogonal to the
validation nuisance tangent.  Hence it is the observed-data canonical
gradient.  This proves Lemma 4.
\end{proof}

\subsection{Proof of Theorem 3}

\begin{proof}
Here the missing full-data component is \(H=\Delta\).  Proposition 3
gives
\[
 \phi_F
 =
 I_0^{-1}
 \{\Delta c_0(U,X)-G_0(U,X)\}.
\]
Because \(G_0(U,X)\) and \(c_0(U,X)\) are \(V\)-measurable,
\[
\begin{aligned}
 E(\phi_F\mid V)
 &=
 I_0^{-1}
 \{E(\Delta\mid V)c_0(U,X)-G_0(U,X)\}\\
 &=
 I_0^{-1}
 \{p_0(V)c_0(U,X)-G_0(U,X)\}.
\end{aligned}
\tag{A.55}
\]
Insert (A.55) into Lemma 4:
\[
\begin{aligned}
 \phi_{\mathrm{eff}}
 &=
 I_0^{-1}\{p_0c_0-G_0\}\\
 &\quad+
 \frac{R}{\pi_0}
 I_0^{-1}\{(\Delta-p_0)c_0\}\\
 &=
 I_0^{-1}
 \left[
 \left\{
 p_0+\frac{R}{\pi_0}(\Delta-p_0)
 \right\}c_0-G_0
 \right]\\
 &=
 I_0^{-1}\{D_0c_0-G_0\}.
\end{aligned}
\]
This is (25)--(26).
\end{proof}

\subsection{Proof of Theorem 4}

\begin{proof}
Theorem 2 gives
\[
 \sqrt n(\widehat\beta-\beta_0)
 =
 \frac1{\sqrt n}\sum_{i=1}^n
 I_0^{-1}\xi_0(O_i)+o_p(1).
\tag{A.56}
\]
Theorem 3 establishes that
\(I_0^{-1}\xi_0=\phi_{\mathrm{eff}}\), the observed-data canonical
gradient.

To verify regularity, let \(\{P_\epsilon:\epsilon\in(-\delta,\delta)\}\)
be an arbitrary regular one-dimensional parametric submodel through
\(P_0\), with score \(S(O)\), and consider local alternatives
\(P_{t/\sqrt n}\).  By the local-uniformity assumption in Theorem 4,
the expansion (A.56) continues to hold with the corresponding local
parameter \(\beta(P_{t/\sqrt n})\) and influence function converging in
\(L_2(P_0)\) to \(\phi_{\mathrm{eff}}\).  Pathwise differentiability
with canonical gradient \(\phi_{\mathrm{eff}}\) gives
\[
 \sqrt n\{\beta(P_{t/\sqrt n})-\beta(P_0)\}
 \longrightarrow
 t\,E\{\phi_{\mathrm{eff}}(O)S(O)\}.
\tag{A.57}
\]
Le Cam's third lemma \citep{bickel1993,tsiatis2006} applied to
\(n^{-1/2}\sum_i\phi_{\mathrm{eff}}(O_i)\) then yields, under
\(P_{t/\sqrt n}^{\,n}\),
\[
 \sqrt n\{\widehat\beta-\beta(P_{t/\sqrt n})\}
 \Rightarrow N_p(0,V_{\mathrm{eff}}),
\]
with a limit distribution independent of \(t\).  This is regularity.

The convolution/information bound for regular estimators in a
semiparametric model \citep{bickel1993,tsiatis2006} states that the covariance of the canonical
gradient is the minimal covariance in Loewner order.  Since OVAC has
that canonical gradient,
\[
 \operatorname{AVar}(\widehat\beta)
 =
 E(\phi_{\mathrm{eff}}\phi_{\mathrm{eff}}^\top)
 =
 V_{\mathrm{eff}}.
\]
Thus OVAC attains the observed-data semiparametric efficiency bound.
\end{proof}

\subsection{Proof of Proposition 4}

\begin{proof}
Let
\[
 h_0(V)=E(\phi_F\mid V),\qquad
 \epsilon=\phi_F-h_0(V),
\]
so \(E(\epsilon\mid V)=0\).  For arbitrary square-integrable \(h(V)\),
write \(d(V)=h(V)-h_0(V)\).  Then
\[
\begin{aligned}
 \phi_h-h_0
 &=
 d+\frac{R}{\pi_0}\{\epsilon-d\}\\
 &=
 \frac{R}{\pi_0}\epsilon
 +
 \left(1-\frac{R}{\pi_0}\right)d.
\end{aligned}
\tag{A.58}
\]
Conditional on \(V\), this has mean zero.  Since
\(R\perp\phi_F\mid V\), the two terms in (A.58) have zero conditional
cross-covariance:
\[
 E\!\left[
 \frac{R}{\pi_0}\epsilon
 \left(1-\frac{R}{\pi_0}\right)d^\top
 \,\middle|\,V
 \right]
 =
 E(\epsilon\mid V)\,
 E\!\left[
 \frac{R}{\pi_0}
 \left(1-\frac{R}{\pi_0}\right)
 \,\middle|\,V
 \right]d^\top
 =0.
\]
Moreover,
\[
 E\!\left[
 \frac{R^2}{\pi_0^2}\epsilon\epsilon^\top
 \,\middle|\,V
 \right]
 =
 \frac{1}{\pi_0}
 \operatorname{Var}(\phi_F\mid V),
\tag{A.59}
\]
and, because \(R^2=R\),
\[
 E\!\left[
 \left(1-\frac{R}{\pi_0}\right)^2
 \,\middle|\,V
 \right]
 =
 \frac1{\pi_0}-1.
\tag{A.60}
\]
Thus
\[
 \operatorname{Var}(\phi_h\mid V)
 =
 \frac{\operatorname{Var}(\phi_F\mid V)}{\pi_0(V)}
 +
 \left\{\frac1{\pi_0(V)}-1\right\}
 d(V)d(V)^\top.
\tag{A.61}
\]
Also \(E(\phi_h\mid V)=h_0(V)\), independent of \(h\).  By the law of
total variance,
\[
\begin{aligned}
 \operatorname{Var}(\phi_h)
 &=
 \operatorname{Var}\{h_0(V)\}
 +
 E\!\left[
 \frac{\operatorname{Var}(\phi_F\mid V)}{\pi_0(V)}
 \right]\\
 &\quad+
 E\!\left[
 \left\{\frac1{\pi_0(V)}-1\right\}
 d(V)d(V)^\top
 \right].
\end{aligned}
\tag{A.62}
\]
The last term is positive semidefinite and is zero exactly when
\(h=h_0\) almost surely on \(\{\pi_0<1\}\).  This proves Loewner
optimality.

For the unaugmented Horvitz--Thompson influence take \(h=0\), hence
\(d=-h_0\).  Projecting the covariance difference in (A.62) onto a
fixed vector \(a\) gives
\[
 E\!\left[
 \left\{\frac1{\pi_0(V)}-1\right\}
 \{a^\top h_0(V)\}^2
 \right]\ge0,
\]
which is (28).
\end{proof}

\subsection{Proof of Proposition 5}

\begin{proof}
We prove consistency in four steps.

\emph{Step 1: the signed cumulative score-linearization measure.}
Write
\[
 H_n^D(t)
 =
 P_n\{\widetilde D\,I(U\le t)\},
 \qquad
 H_0^D(t)
 =
 E\{\Delta I(U\le t)\}.
\]
By (A.2),
\(E\{D_0I(U\le t)\}=H_0^D(t)\).  Moreover,
\[
 \sup_{t\le\tau}
 |H_n^D(t)-P_n\{D_0I(U\le t)\}|
 \le P_n|\widetilde D-D_0|
 =o_p(1).
\tag{A.63}
\]
The class
\(\{D_0I(U\le t):t\le\tau\}\) is Glivenko--Cantelli because \(D_0\) is
bounded and the indicator class is VC.  Hence
\[
 \sup_{t\le\tau}|H_n^D(t)-H_0^D(t)|=o_p(1).
\tag{A.64}
\]
The Cox compensator identity gives
\[
 dH_0^D(t)=E\{dN(t)\}=s_0(t)d\Lambda_0(t).
\tag{A.65}
\]

The estimator in (29) can be written exactly as the Stieltjes integral
\[
 \widehat A_D(t)
 =
 \int_0^t
 \frac{1}{S_n^{(0)}(u,\widehat\beta)}
 \,dH_n^D(u).
\]
Likewise, by (A.65),
\[
 \Lambda_0(t)
 =
 \int_0^t\frac1{s_0(u)}\,dH_0^D(u).
\]
Set
\(q_n(u)=1/S_n^{(0)}(u,\widehat\beta)\),
\(q_0(u)=1/s_0(u)\), and
\(K_n=H_n^D-H_0^D\).  Then
\[
 \widehat A_D(t)-\Lambda_0(t)
 =
 \int_0^t(q_n-q_0)\,dH_n^D
 +
 \int_0^t q_0\,dK_n.
\tag{A.66}
\]
The first term satisfies
\[
 \sup_{t\le\tau}
 \left|
 \int_0^t(q_n-q_0)\,dH_n^D
 \right|
 \le
 \|q_n-q_0\|_\infty\,TV(H_n^D)
 =o_p(1),
\]
because \(\|q_n-q_0\|_\infty=o_p(1)\) and
\(TV(H_n^D)\le P_n|\widetilde D|=O_p(1)\).
For the second term, \(s_0(t)\) is nonincreasing in \(t\), bounded away
from zero, so \(q_0=1/s_0\) is bounded and of bounded variation.
Integration by parts gives, uniformly in \(t\),
\[
 \left|\int_0^tq_0\,dK_n\right|
 \le
 \|K_n\|_\infty
 \{2\|q_0\|_\infty+TV(q_0)\}
 =o_p(1)
\]
by (A.64).  Therefore
\[
 \sup_{t\le\tau}
 |\widehat A_D(t)-\Lambda_0(t)|=o_p(1).
\tag{A.67}
\]
Also
\[
 TV(\widehat A_D)
 \le
 \|q_n\|_\infty TV(H_n^D)
 =O_p(1).
\tag{A.68}
\]

\emph{Step 2: consistency of the estimated risk-set contribution.}
For each \(i\), write
\[
 \widehat w_i=e^{\widehat\beta^\top X_i},\qquad
 w_{0i}=e^{\beta_0^\top X_i}.
\]
Adding and subtracting intermediate terms gives
\[
\begin{aligned}
 \widehat G_i-G_{0i}
 &=
 (\widehat w_i-w_{0i})
 \int_0^{U_i}\widehat c_i\,d\widehat A_D\\
 &\quad+
 w_{0i}\int_0^{U_i}(\widehat c_i-c_{0i})\,d\widehat A_D\\
 &\quad+
 w_{0i}\int_0^{U_i}c_{0i}\,d(\widehat A_D-\Lambda_0).
\end{aligned}
\tag{A.69}
\]
The first term is \(o_p(1)\) uniformly in \(i\) because bounded \(X\)
implies
\(\sup_i|\widehat w_i-w_{0i}|=o_p(1)\), while
\(\sup_i\|\widehat c_i\|=O_p(1)\) and
\(TV(\widehat A_D)=O_p(1)\).  The second is uniformly \(o_p(1)\)
because
\[
 \sup_{i,t}
 \|\widehat c_i(t)-c_0(t,X_i)\|
 =
 \sup_t\|\bar{X}_n(t,\widehat\beta)-\bar{X}_0(t)\|
 =o_p(1).
\]
For the third term, \(c_0(t,X_i)=X_i-\bar{X}_0(t)\) has total variation
equal to that of \(\bar{X}_0\), uniformly in \(i\).  Integration by parts
therefore yields
\[
 \sup_i
 \left|
 \int_0^{U_i}c_{0i}\,d(\widehat A_D-\Lambda_0)
 \right|
 \le
 C\,
 \|\widehat A_D-\Lambda_0\|_\infty
 =o_p(1).
\]
Thus
\[
 \max_{1\le i\le n}\|\widehat G_i-G_{0i}\|=o_p(1),
 \qquad
 P_n\|\widehat G-G_0\|^2=o_p(1).
\tag{A.70}
\]

\emph{Step 3: consistency of the estimated influence values.}
From (A.18)--(A.19),
\[
 P_n|\widetilde D-D_0|^2=o_p(1),
\tag{A.71}
\]
and uniform risk-set convergence gives
\[
 P_n\|
 \widetilde D\,\widehat c(U)-D_0c_0(U,X)
 \|^2=o_p(1).
\tag{A.72}
\]
The same decomposition used in (A.39) gives
\(\widehat I\to_p I_0\), hence
\(\widehat I^{-1}\to_p I_0^{-1}\).  Combining (A.70)--(A.72),
\[
 P_n\|
 \widehat\phi-\phi_{\mathrm{eff}}
 \|^2=o_p(1).
\tag{A.73}
\]

\emph{Step 4: sample covariance.}
By Cauchy--Schwarz and (A.73),
\[
 P_n(\widehat\phi\widehat\phi^\top)
 -
 P_n(\phi_{\mathrm{eff}}\phi_{\mathrm{eff}}^\top)
 =o_p(1),
\qquad
 \bar{\widehat\phi}-P_n\phi_{\mathrm{eff}}=o_p(1).
\]
The law of large numbers and \(E\phi_{\mathrm{eff}}=0\) imply
\[
 P_n(\phi_{\mathrm{eff}}\phi_{\mathrm{eff}}^\top)
 \to_p V_{\mathrm{eff}},
 \qquad
 P_n\phi_{\mathrm{eff}}\to_p0.
\]
Multiplying (32) by \(n\) gives the empirical covariance of the
\(\widehat\phi_i\)'s with divisor \(n\).  Therefore
\[
 n\,\widehat{Var}(\widehat\beta)\to_pV_{\mathrm{eff}},
\]
as claimed.
\end{proof}

\subsection{Proof of the validation-variance decomposition and Theorem 5}

\begin{proof}
Let \(\theta=a^\top\beta_0\),
\[
 Z=a^\top\phi_F,\qquad
 m_a(V)=E(Z\mid V),\qquad
 \sigma_a^2(V)=\operatorname{Var}(Z\mid V).
\]
Under a design with validation probability \(\pi(V)\), Lemma 4 gives
\[
 Z_{\mathrm{eff}}
 =
 m_a(V)+\frac{R}{\pi(V)}\{Z-m_a(V)\}.
\tag{A.74}
\]
Conditional on \(V\),
\[
 E(Z_{\mathrm{eff}}\mid V)=m_a(V)
\]
and, using \(R\perp Z\mid V\),
\[
 \operatorname{Var}(Z_{\mathrm{eff}}\mid V)
 =
 \frac{\sigma_a^2(V)}{\pi(V)}.
\tag{A.75}
\]
The law of total variance yields
\[
 \operatorname{Var}(Z_{\mathrm{eff}})
 =
 \operatorname{Var}\{m_a(V)\}
 +
 E\!\left\{\frac{\sigma_a^2(V)}{\pi(V)}\right\},
\]
which proves \eqref{eq:35}.  Only the second term depends on the validation rule.

We therefore minimize the convex functional
\[
 J(\pi)=E\!\left\{\frac{\sigma_a^2(V)}{\pi(V)}\right\}
\tag{A.76}
\]
over the convex set
\[
 \mathcal P_q=
 \{\pi:\ E\pi=q,\ \pi_{\min}\le\pi\le1\}.
\]
For \(\sigma_a(V)>0\), the map
\(x\mapsto\sigma_a^2(V)/x\) is strictly convex on \(x>0\).  Introduce a
multiplier \(\lambda\) for the budget and pointwise KKT multipliers
\(\alpha(V)\ge0\) and \(\gamma(V)\ge0\) for the lower and upper bounds.
The pointwise Lagrangian is
\[
 \ell_V(x)
 =
 \frac{\sigma_a^2(V)}{x}
 +\lambda x
 +\alpha(V)\{\pi_{\min}-x\}
 +\gamma(V)\{x-1\}.
\tag{A.77}
\]
Stationarity is
\[
 -\frac{\sigma_a^2(V)}{x^2}
 +\lambda-\alpha(V)+\gamma(V)=0,
\tag{A.78}
\]
with complementary slackness
\[
 \alpha(V)\{x-\pi_{\min}\}=0,\qquad
 \gamma(V)(1-x)=0.
\tag{A.79}
\]
If neither bound is active, \(\alpha=\gamma=0\) and
\[
 x=\frac{\sigma_a(V)}{\sqrt\lambda}
 =\kappa\sigma_a(V),
 \qquad
 \kappa=\lambda^{-1/2}.
\tag{A.80}
\]
If (A.80) falls below \(\pi_{\min}\), the lower-bound KKT condition
forces \(x=\pi_{\min}\); if it exceeds \(1\), the upper-bound condition
forces \(x=1\).  Hence on \(\{\sigma_a>0\}\)
\[
 \pi_{\mathrm{opt}}(V)
 =
 \min\left[
 1,\max\{\pi_{\min},\kappa\sigma_a(V)\}
 \right].
\tag{A.81}
\]
Because the objective and constraints are convex and Slater's condition
holds for the interior budget \(q\in(\pi_{\min},1)\), the KKT
conditions are sufficient for global optimality.

When \(\Pr(\sigma_a>0)=1\), define
\[
 F(\kappa)
 =
 E\left[
 \min\{1,\max(\pi_{\min},\kappa\sigma_a(V))\}
 \right].
\]
The function is continuous and nondecreasing in \(\kappa\), with
\(F(0)=\pi_{\min}\) and
\(\lim_{\kappa\to\infty}F(\kappa)=1\) by dominated convergence.
Therefore for every \(q\in(\pi_{\min},1)\) there exists a
\(\kappa>0\) satisfying \(F(\kappa)=q\).  Strict convexity on the
positive-\(\sigma_a\) region makes the minimizing allocation unique
there, up to regions forced to the same boundary.

If \(\sigma_a=0\) on a positive-probability set, the contribution
\(\sigma_a^2/\pi\) is identically zero there.  Thus the allocation on
that set is irrelevant to \(J(\pi)\); any residual budget after the
optimal clipped allocation on \(\{\sigma_a>0\}\) can be placed there
without changing the objective.  This proves the stated zero-variance
qualification.

If both pointwise bounds are inactive, (A.80) holds everywhere and the
budget equation gives
\[
 q=E\pi_{\mathrm{opt}}
   =\kappa E\sigma_a,
 \qquad
 \kappa=\frac{q}{E\sigma_a},
\]
hence
\[
 \pi_{\mathrm{opt}}(V)
 =
 \frac{q\,\sigma_a(V)}{E\{\sigma_a(V)\}}.
\]
For the endpoint budgets, the only feasible constant-bound solutions are
\(\pi\equiv\pi_{\min}\) when \(q=\pi_{\min}\) and
\(\pi\equiv1\) when \(q=1\), completing the proof.
\end{proof}

\subsection{Derivation of the contrast-specific conditional variance and proof of Corollary 2}

\begin{proof}
From Proposition 3,
\[
 a^\top\phi_F
 =
 a^\top I_0^{-1}
 \{\Delta c_0(U,X)-G_0(U,X)\}.
\]
Conditional on \(V\), the quantities \(U,X,c_0,G_0\) are fixed and only
\(\Delta\) is random.  Therefore
\[
\begin{aligned}
 \sigma_a^2(V)
 &=
 \operatorname{Var}(a^\top\phi_F\mid V)\\
 &=
 \{a^\top I_0^{-1}c_0(U,X)\}^2
 \operatorname{Var}(\Delta\mid V)\\
 &=
 \{a^\top I_0^{-1}c_0(U,X)\}^2
 p_0(V)\{1-p_0(V)\},
\end{aligned}
\]
which is \eqref{eq:34}.

For \(\theta=\beta_{0j}\), take \(a=e_j\).  Then
\[
 \sigma_a(V)
 =
 |e_j^\top I_0^{-1}c_0(U,X)|
 \sqrt{p_0(V)\{1-p_0(V)\}}.
\tag{A.82}
\]
The clipped optimal rule in (36) is monotone in \(\sigma_a(V)\).
For fixed Cox leverage
\(|e_j^\top I_0^{-1}c_0(U,X)|\), the Bernoulli standard deviation
\(\sqrt{p(1-p)}\) is maximized at \(p=1/2\).  For fixed classification
uncertainty, the allocation increases with the absolute Cox leverage.
Thus the design preferentially validates records that are simultaneously
classification-ambiguous and influential for the target coefficient,
as stated in Corollary 2.
\end{proof}

\section{Technical conditions and empirical-process
details}

This appendix records sufficient empirical-process conditions used
explicitly in the proofs above. They are not claimed to be minimal. Let \(B\) be an
open, relatively compact neighborhood of \(\beta_{0}\). Assume the map
\(\beta \mapsto exp\left( \beta^{\top}X \right)\) is dominated by an
integrable envelope on \(B\), and
\(\inf_{t \leq \tau,\,\beta \in B}s^{(0)}(t,\beta) > c > 0\). The
classes
\(\left( Y(t)\exp\left( \beta^{\top}X \right)X^{\otimes k}:t \leq \tau,\beta \in B \right)\),
\(k = 0,1,2\), are Glivenko-Cantelli and admit the usual stochastic
equicontinuity required for the Cox score. Bounded \(X\) is a convenient
sufficient condition.  For Proposition 5, assume additionally that
\(t\mapsto\bar{X}_0(t)\) has bounded variation on \([0,\tau]\); this
ensures the Stieltjes integration-by-parts bound used to transfer
uniform convergence of the signed cumulative measure to \(G_0\).

For the validation nuisances, define \(L_{2}\) norms under the phase-I
distribution of \(V\). Cross-fitting partitions the data into \(K\)
fixed folds. Conditional on the training data for fold \(k\), the
held-out pseudo-events are independent draws from the evaluation
distribution with nuisance functions fixed. If \({\widehat{p}}_{k}\) and
\({\widehat{\pi}}_{k}\) are uniformly bounded,
\({\widehat{\pi}}_{k} \geq \varepsilon/2\), and their \(L_{2}\) errors
converge to zero, the variance of the centered score difference is
\(o_{p}(1)\) after multiplying by the nuisance error. Summing across
finitely many folds preserves the rate.

The product-rate remainder follows exactly from Proposition 1 and
therefore does not require entropy conditions. The only
empirical-process complexity arises from replacing the population
risk-set mean \(\bar{X}\) by \(\bar{X}_{n}\) and from estimating
\(\beta\). Those are the same objects appearing in ordinary Cox
regression and can be handled using the standard
martingale/empirical-process expansion. This separation is one reason
the present endpoint-only problem is theoretically attractive.

If \(\pi_{0}\) is estimated rather than known, cross-fitting can be used for
\(\pi\) as well. If the validation design is deterministic within strata with fixed
phase-II sample sizes, the Bernoulli formulation can be replaced by
finite-population two-phase sampling. The leading augmentation identity
continues to hold with design expectations, but the asymptotic variance
must reflect sampling without replacement. Breslow and Wellner (2007)
and Saegusa and Wellner (2013) provide the required weighted
empirical-process machinery. The main text uses Bernoulli validation for
clarity.

The asymptotic efficiency statement assumes the observed-data model in
which the conditional law of \(\Delta\) given \(V\) is otherwise
unrestricted and validation is coarsening at random. If additional
parametric structure is imposed on the misclassification mechanism, a
smaller semiparametric model could in principle have a lower efficiency
bound. OVAC is efficient relative to the stated nonparametric bridge
model, not relative to every more restrictive model one might impose.

\section{Computational
details}

\subsection{Risk-set calculation}

For time-fixed covariates and no tied observed times, sort records in
descending \(U\). At a candidate \(\beta\), cumulative sums of
\(\exp\left( \beta^{\top}X \right)\),
\(\exp\left( \beta^{\top}X \right)X\), and
\(\exp\left( \beta^{\top}X \right)XX^{\top}\) generate \(S_{n}^{(0)}\),
\(S_{n}^{(1)}\), and \(S_{n}^{(2)}\) for every subject in
\(O\left( np^{2} + n\log n \right)\) time. The score and information
then require a single pass through \(\widetilde{D}\). With ties, use a
common risk-set value for all records at the same observed time and the
usual Breslow convention.

\subsection{Cross-fitting
details}

When validation is sparse, folds should be stratified by \(R\) and, if
feasible, \(\widetilde{\Delta}\) so that each training split contains
enough validated apparent events and non-events to fit the bridge. A
\(K\) of 5 is a reasonable default; larger \(K\) increases training size
but can destabilize the fold-specific validation composition.
Predictions should be stored for every observation and combined before
solving the Cox score. The final Cox fit is not repeated fold by fold.

\subsection{Pseudocode}

\begin{enumerate}
\def\labelenumi{\arabic{enumi}.}
\tightlist
\item
  Input \(U\), \(X\), \(\widetilde{\Delta}\), auxiliary variables,
  validation indicator \(R\), gold-standard \(\Delta\) when \(R = 1\),
  and known or estimable \(\pi\).
\item
  Split subjects into \(K\) folds.
\item
  For each held-out fold, fit \(p_{k}(V) = \Pr(\Delta = 1 \mid V)\) on
  validated training records and, when needed, fit
  \(\pi_{k}(V) = \Pr(R = 1 \mid V)\) on all training records.
\item
  Predict \({\widehat{p}}_{i}\) and \({\widehat{\pi}}_{i}\) for the
  held-out records and form
  \({\widetilde{D}}_{i} = {\widehat{p}}_{i} + R_{i}\left( \Delta_{i} - {\widehat{p}}_{i} \right)/{\widehat{\pi}}_{i}\).
\item
  Using all subjects, solve
  \(\sum_{i}^{}{\widetilde{D}}_{i}\left( X_{i} - \bar{X}_{n}\left( U_{i};\beta \right) \right) = 0\)
  by damped Newton-Raphson.
\item
  Estimate the score-linearization influence quantities and standard
  errors under the intersection model.
\item
  Return \(\widehat{\beta}\), its covariance estimate, nuisance
  diagnostics, and convergence information.
\end{enumerate}

\subsection{Numerical safeguards}

Validation weights should not be truncated casually when \(\pi_{0}\) is
known by design, because truncation changes the estimating equation. If
estimated \(\widehat{\pi}\) becomes extremely small because of model
extrapolation, the primary remedy is better overlap or a better
validation design. For exploratory analyses, prespecified truncation can
be used with a sensitivity analysis, but the resulting target should be
treated as an approximation.

Pseudo-events can be negative, so software that assumes a binary event
status cannot always be used as a black box. The score itself remains
smooth. A damped Newton update with step-halving can enforce monotone
reduction of \(\parallel U_{n}(\beta) \parallel\). Convergence should be
declared only when both the score norm and parameter increment are
small.

For analytic variance estimation, the default implementation uses the
direct pseudo-event score-linearization measure in (29). A bridge-based
version obtained by replacing \(\widetilde{D}\) with \(\widehat{p}\) is
retained only as an intersection-model diagnostic because it is
asymptotically equivalent when \(\widehat{p}\) is consistent. In finite
samples the two can differ. The Monte Carlo calibration in Section 9
reports mean analytic SE, empirical SD, their ratio, and 95\% coverage.
Under one-sided robustness with an estimated active nuisance model, use
a nuisance-refitting subject-level bootstrap unless the corresponding
nuisance influence term has been derived.

\section{Additional derivations and
relationships}

\subsection{Closed-form nuisance
remainder}

Equation~\eqref{eq:16} can be rewritten in terms of relative validation-model
error. Define

\[r_{\pi}(V) = \frac{\pi(V) - \pi_{0}(V)}{\pi(V)}.\]

Then

\[
\Psi(\beta;p,\pi) = \Psi_{F}(\beta) + E\left\lbrack c_{\beta}(U,X)r_{\pi}(V)\left( p(V) - p_{0}(V) \right) \right\rbrack.
\tag{D.1}\label{eq:D1}
\]

This form makes the role of positivity explicit. Even a moderate
absolute error in \(\pi\) can be magnified where validation
probabilities are small.

\subsection{Known sensitivity and
specificity}

Suppose \(\widetilde{\Delta}\) has nondifferential sensitivity \(Se\)
and specificity \(Sp\) conditional on phase-I variables \(V_{0}\) that
exclude \(\widetilde{\Delta}\). Let
\(q\left( V_{0} \right) = Pr\left( \Delta = 1 \mid V_{0} \right)\).
Bayes' rule gives

\[
\Pr\left( \Delta = 1 \mid \widetilde{\Delta} = 1,V_{0} \right) = \frac{Se\, q\left( V_{0} \right)}{Se\, q\left( V_{0} \right) + (1 - Sp)\left( 1 - q\left( V_{0} \right) \right)},
\tag{D.2}\label{eq:D2}
\]

and

\[
\Pr\left( \Delta = 1 \mid \widetilde{\Delta} = 0,V_{0} \right) = \frac{(1 - Se)q\left( V_{0} \right)}{(1 - Se)q\left( V_{0} \right) + Sp\left( 1 - q\left( V_{0} \right) \right)}.
\tag{D.3}\label{eq:D3}
\]

OVAC does not require this parameterization; it can be useful when
external information makes \(Se\) and \(Sp\) credible and transportable.

\subsection{Algebraic relationship to Liu and Wang
(2010)}

The connection to earlier AIPW Cox estimation is exact at the
event-contribution level. Let \(\xi\) denote the indicator that the true
failure status is observed, let \(m(W)\) be a working conditional mean,
and let \(\pi(W)\) be the observation probability. The augmented term

\[\frac{\xi\Delta}{\pi(W)} + \left( 1 - \frac{\xi}{\pi(W)} \right)m(W)\]

can be rearranged as

\[
m(W) + \frac{\xi}{\pi(W)}\left( \Delta - m(W) \right).
\tag{D.4}\label{eq:D4}
\]

Equation~\eqref{eq:D4} is the same pseudo-event map as equation~\eqref{eq:9}.
Consequently, OVAC should not be interpreted as replacing the classical
AIPW estimator with a different point-estimating equation. Its
methodological role is to provide a cross-fitted orthogonal formulation,
an explicit product-form drift calculation, the first-order empirical
risk-set linearization needed for analytic inference, and a direct link
to modern validation design. This relationship distinguishes extension
from reinvention while making the additional inferential structure
explicit.

\subsection{Relationship to generalized raking and adaptive
validation}

Generalized raking uses phase-I auxiliary variables to calibrate
validation weights so that weighted phase-II totals reproduce phase-I
information. Semiparametrically, the most informative auxiliary
quantities approximate conditional expectations of complete-data
influence functions. In the endpoint-only setting, the hidden component
of the Cox influence function is particularly simple because only
\(\Delta\) is missing conditional on \(V\). The bridge \(p_{0}(V)\)
therefore provides a direct representation of the relevant conditional
expectation.

This specialization also clarifies the relationship to optimal
validation designs developed for error-prone survival data. General
two-phase design theory allocates validation where the conditional
variance of the complete-data influence function is large. Equation (34)
reduces that criterion to the product of a Cox leverage term and the
Bernoulli uncertainty \(p_{0}(V)\left( 1 - p_{0}(V) \right)\). The
closed form is useful computationally, but it is a specialization of the
broader influence-function design principle rather than a claim that
optimal two-phase validation is new.

\subsection{Rare-event behavior}

When true events are rare, \(p_{0}(V)\) is small for most records and
the conditional Bernoulli variance
\(p_{0}(V)\left( 1 - p_{0}(V) \right)\) is correspondingly small. The
records that dominate the design criterion are those for which phase-I
information raises event uncertainty above the background rate and for
which the Cox contrast is influential. This explains why uniform
validation can be inefficient and why oversampling apparent cases alone
need not be optimal: an obvious apparent case may have little
classification uncertainty, whereas an ambiguous record with a large
influence contribution may be more valuable.

\end{document}